\documentclass[10pt]{article}

\usepackage{amsmath}
\usepackage{amssymb}
\usepackage{textcomp}
\usepackage{graphicx}
\usepackage{cite}
\usepackage{color} 
\usepackage{setspace} 
\usepackage{ifthen}  % Conditional 
\usepackage{multicol}   %Columns
\usepackage[utf8]{inputenc} %unicode support
\usepackage[hang,flushmargin]{footmisc}
\usepackage[greek,english]{babel}
\usepackage{sectsty}  %% change section title fonts
\allsectionsfont{\usefont{OT1}{phv}{b}{n}\normalsize\selectfont}

   \def\url#1{}

\newboolean{publ}

\usepackage[labelfont=bf,labelsep=period,justification=raggedright]{caption}

\makeatletter
\renewcommand{\@biblabel}[1]{\quad#1.}
\makeatother

\date{}

\newenvironment{erivesformat}{\fussy\setboolean{publ}{true}}{\fussy}

\newcommand{\SOPH}{S\textsc{ophophora}}               %% Sub-genus capitalized, then small caps
\newcommand{\Dros}{{\it Drosophila}}
\newcommand{\mel}{{\it D. melanogaster}}
\newcommand{\ana}{{\it D. ananassae}}
\newcommand{\pse}{{\it D. pseudoobscura}}
\newcommand{\wil}{{\it D. willistoni}}
\newcommand{\vir}{{\it D. virilis}}
\newcommand{\vn}{{\it vn}}                            %% italicized gene abbreviations
\newcommand{\vnd}{{\it vnd}}
\newcommand{\brk}{{\it brk}}
\newcommand{\sog}{{\it sog}}
\newcommand{\rhomb}{{\it rho}}
\newcommand{\Dl}{{\it Delta}}                         %% no abbrev to avoid confusion with Dorsal
\newcommand{\eg}{{e.g.}}                              %% CMOS: Common Latin words should not be italicized. 
\newcommand{\ie}{{i.e.}}                              %% CMOS: "e.g." and "i.e." should not be italicized.
\newcommand{\insitu}{\textit{in~situ}}
\newcommand{\cis}{\textit{cis}}
\newcommand{\trans}{{\it trans}}
\newcommand{\fiveprime}[1]{$5^{\prime}$-\texttt{#1}}  %% Use for all DNA sequence
\newcommand{\DNA}[1]{\texttt{#1}}                     %% Use for DNA sequence where 5' label is cumbersome

\newcommand{\hS}{\hspace{-0.65pt}}                    %% Use for adjusting tighter tracking in cis-element names
\newcommand{\hs}{\hspace{-0.18pt}}                    %% Use for adjusting tighter tracking in cis-element names
\newcommand{\Da}{\mbox{\textit{D}\hs$\alpha$}}
\newcommand{\Db}{\mbox{\textit{D}\hs$\beta$}}

\newcommand{\E}[1]{\mbox{\textit{E\hS{}(\hs\DNA{#1}\hS)}}}
\newcommand{\ECAT}{\mbox{\E{CA}\hs\textit{\DNA{T}}}}
\newcommand{\NDb}{\mbox{\textit{N-D\hs}$\beta$}}
\newcommand{\NECAT}{\mbox{\textit{N-\ECAT}}}
\newcommand{\SUH}{\mbox{\textit{S\hS{}U\hS{}H}}}
\newcommand{\NEE}[1]{NEE$_{#1}$}                     %% Takes gene locus as argument
\newcommand{\EtoD}{\textit{E-to-D}}
\newcommand{\hEtoD}{\textit{\textbf{\EtoD}}}
\newcommand{\hDros}{\textit{\textbf{Drosophila}}}    %% Header version
            
\newcommand{\hcis}{\textit{\textbf{cis}}}  
\newcommand{\hmel}{\textit{\textbf{D. melanogaster}}}
\newcommand{\hwil}{\textit{\textbf{D. willistoni}}}
\newcommand{\hana}{\textit{\textbf{D. ananassae}}}
\newcommand{\hvnd}{\textit{\textbf{vnd}}}
\newcommand{\hDl}{\textit{\textbf{Delta}}}
\newcommand{\hDa}{\textit{\textbf{\Da}}}
\newcommand{\hDb}{\textit{\textbf{\Db}}}
\newcommand{\hNDb}{\textit{\textbf{\NDb}}}
\newcommand{\hECAT}{\textit{\textbf{\ECAT}}}

\newcommand{\hNEE}[1]{\textbf{\NEE{#1}}}
\newcommand{\hDNA}[1]{\textbf{\DNA{#1}}}
\newcommand{\term}[1]{\textit{#1}}               %% Use this style for all referring to the name of a term, or first use thereof
\newcommand{\Xomit}[1]{}  % Use to omit or X-out text.
\begin{document}
\begin{erivesformat}
\begin{centering}
{\Large \textbf{
\usefont{OT1}{phv}{b}{n}\selectfont
Dynamic evolution of precise regulatory encodings creates the clustered signature of developmental enhancers}
} \\

\vskip 0.5cm 

{\usefont{OT1}{phv}{b}{n}\selectfont Albert Erives$^{1\ast}$ and Justin Crocker$^{1}$} \\
$^1$ \textit{Dept. of Biological Sciences, Dartmouth College, Hanover, NH, U.S.A.}  \\
$^\ast$ E-mail: Albert.J.Erives@Dartmouth.edu \\
(Dated: April 7th, 2010)\let\thefootnote\relax\footnotetext{
 \copyright{} 2010 Erives and Crocker. This is an open-access article
 distributed under the terms of the Creative Commons Attribution License, which
 permits unrestricted use, distribution, and reproduction in any medium,
 provided the original authors are credited. This article is archived in
 the Quantitative Biology section of arXiv.org, which is maintained by Cornell
 University Library, and was first presented in a platform session at the 51st Annual Drosophila
 Research Conference, April 8th, 2010, in Washington, DC, USA.
}\\

\end{centering}

\vskip 0.5cm 

{\small\usefont{OT1}{phv}{b}{n}\selectfont \noindent A morphogenic protein
known as Dorsal patterns the embryonic dorsoventral body axis of \hDros{} by
binding to transcriptional enhancers across the genome.  Each such enhancer
activates a neighboring gene at a unique threshold concentration of Dorsal.
The presence of Dorsal binding site clusters in these enhancers and of similar
clusters in other enhancers has motivated models of threshold-encoding in site
density. However, we found that the precise length of a spacer separating a
pair of specialized Dorsal and Twist binding sites determines the
threshold-response.  Despite this result, the functional range determined by
this spacer element as well as the role and origin of its surrounding Dorsal
site cluster remained completely unknown.  Here, we experiment with enhancers
from diverse \hDros{} genomes, including the large uncompacted genomes from
\textit{\textbf{ananassae}} and \textit{\textbf{willistoni}}, and report three
major interdependent results.  First, we map the functional range of the
threshold-encoding spacer variable.  Second, we show that the majority of sites
at the cluster are non-functional divergent elements that have been separated
beyond the encoding's functional range.  Third, we verify an evolutionary model
involving the frequent replacement of a threshold encoding, whose precision is
easily outdated by shifting accuracy.  The process by which encodings are
replaced by newer ones is facilitated by the palindromic nature of the Dorsal
and Twist binding motifs and by intrinsic repeat-instability in the specialized
Twist binding site, which critically impacts the length of the spacer linking
it to Dorsal.  Over time, the dynamic process of selective deprecation and
replacement of encodings adds to a growing cluster of deadened elements, or
\textit{\textbf{necro-elements}}, and strongly biases local sequence
composition.  Necro-element plaques are associated with mature enhancers that
are older than 10~My but not with newer lineage-specific enhancers that employ
identical logic.  We conclude that the clustered signature of most enhancers
results from long histories of selective ``maintenance'' of precise encodings
via facile deprecation and equally facile replacement.}

\ifthenelse{\boolean{publ}}{\begin{multicols}{2}}{}
\vskip -0.5cm
\section*{Introduction} 
\begin{quote}
Nothing Gold Can Stay\\
\\
Nature's first green is gold,\\
Her hardest hue to hold.\\
Her early leaf's a flower;\\
But only so an hour.\\
Then leaf subsides to leaf.\\
So Eden sank to grief,\\
So dawn goes down to day,\\
Nothing gold can stay.\\
\\
---Robert Frost, {\it New Hampshire} (1923)
\end{quote}

\noindent How genetic information is encoded in DNA is a central question in
biology.  In many cases, natural selection acts efficaciously on regulatory DNA
sequences, which specify the precise conditions under which a gene product is
made by a cell \cite{Prudhomme:2007qp, Carroll:2008tk, Wang:2002wj,
Wittkopp:2004sy, Marcellini:2006by, McGregor:2007lt, Crocker:2008pb,
Prabhakar:2008vf, Williams:2008eu, Wittkopp:2008oq, Shirangi:2009hb}. However,
unlike the precise protein-encoding scheme, few general principles have emerged
for regulatory encoding.  The identification of such principles would
facilitate understanding of genomic regulatory DNAs and advance many areas of
biological investigation. 

One general feature of regulatory DNAs, which include the transcriptional
\term{enhancers}, is the use of combinatorial codes of transcription factor
(TF) binding sites \cite{Arnone:1997yk}.  This feature allows an enhancer to
activate its gene only if it binds a specific combination of different TF
proteins.  A less understood general feature is the clustering of multiple
binding sites for a single TF operating at an enhancer \cite{Wunderlich:2009mz}.
This unexplained cluster signature has motivated several bioinformatic screens
that exploit binding site density to identify functional enhancers
\cite{Berman:2002dz, Markstein:2002dz}.  Such methods detect both functional
enhancers and non-functional sequences. Moreover, these methods are not yet
predictive of the exact responses encoded by active enhancers bearing site
clusters.

Concentration-specific threshold responses are a property of most regulatory
DNAs that function through recruitment of DNA-binding factors
\cite{Ptashne:2004ft}.  However, developmental enhancers that read classical
morphogen concentration gradients \cite{Wolpert:1989dk} are ideal subjects in
decoding regulatory DNA sequences, and their functional features.  Different
enhancers with variably-dense clusters of binding sites for the same TF are
each responsive to their own unique threshold concentration.  Such DNAs can be
studied comparatively to identify the variables that encode the concentration
threshold setting.  In principle, such a variable might be encoded in one of
several non-exclusive categories: i) the formulaic combination of adjacent
binding sites for TFs acting synergistically; ii) the range of sequences that
determine the affinity or allostery of a DNA-bound TF (\term{functional
grammars}); and iii) the higher-order organizational arrangement of binding
sites (\term{functional syntaxes}).  

Two well-studied systems of morphogen-responsive enhancers are those that read
the Bicoid and Dorsal morphogen concentration gradients that pattern the
anterior/posterior (A/P) and dorsal/ventral (D/V) axes of the \Dros{} embryo,
respectively \cite{Anderson:1985zl, Jiang:1991dq, Small:1991kk, Ip:1992wj,
Norris:1992yg, Reinitz:1995gd, Jaeger:2004ph, Moussian:2005jx, Gregor:2007dn,
Gregor:2007xe, Reinitz:2007pi, Gregor:2008rw}.  Like most enhancers, these DNAs
contain clusters of binding sites, which in this case correspond to those for
Bicoid, Dorsal, and their DNA-binding co-factors. This clustering has prompted
several complex ``cluster code'' models that integrate site number, quality,
and density parameters to determine the threshold readout
\cite{Papatsenko:2005it, Zinzen:2006fk, Janssens:2006la}.  Paradoxically
however, the apparent phenotypic robustness of this ``cluster code'' to
mutational divergence has been taken to mean that this full parameter set is
simultaneously flexible and determinative \cite{Brown:2007uk, Hare:2008fu,
Liberman:2009rp, Cameron:2009mz}.  

To address how concentration-threshold responses are encoded in Dorsal target
enhancers, we asked whether there exist a unique, subset of \term{specialized}
TF binding sites in co-clusters of sites for Dorsal, Twist, and Suppressor of
Hairless [Su(H)] \cite{Erives:2004pd}.  Specialized binding motifs, as
identified across equivalent enhancers present in a genome and across related
lineages, do not manifest the full-range of sequences known to be bound by
these factors and may signify regulatory sub-functionalizations.  With this
approach, we identified two different specialized binding sites for Dorsal, as
well as specialized binding sites for Twist and Su(H) \cite{Erives:2004pd}.
Since then, we have formally referred to DNA sequences that both drive
expression in the lateral embryonic ectoderm and contain this particular
collection of specialized binding sites as \term{Neurogenic Ectodermal
Enhancers}, or NEEs \cite{Erives:2004pd, Crocker:2008pb, Crocker:2008pi}. 

We found that the NEE at the \vnd{} locus, or \NEE{vnd}, is conserved in
\Dros{} and mosquitos. As such it was present in the latest common ancestor of
dipterans $\sim$240--270 million years ago (Mya)
\cite{Grimaldi:2005kx,Bertone:2008se}, or at least $>$200 Mya
\cite{Wiegmann:2009bb}. We found that conserved ``canonical'' NEEs occur at the
\rhomb{}, \vnd{}, \brk{}, and \vn{} loci across the \Dros{} genus
\cite{Crocker:2008pb}.  As such, the canonical NEEs were acquired prior to
\Dros{} diversification over 40 Mya \cite{Crocker:2008pb}. We also found a more
recently evolved member of this enhancer class, \NEE{sog}, in the \sog{} locus
of the melanogaster subgroup, which began diverging $\sim$20 Mya
\cite{Crocker:2008pb}.  Thus, NEE-type regulatory sequences have been evolving
at various unrelated loci within a period spanning the last $\sim$250 My.

NEEs function by recruiting both Dorsal, a {\it rel}-homology domain
(RHD)-containing TF, and its synergistic bHLH co-activator Twist, whose
expression mirrors the Dorsal morphogen gradient \cite{Jiang:1991dq, Ip:1992ff,
Ip:1992ec, Jiang:1993lh, Gonzalez-Crespo:1993fu, Szymanski:1995qf}.  In
addition to having sites for Dorsal and Twist, NEEs possess sites for Su(H) and
Snail. Su(H) is a highly-conserved TF that mediates transcriptional responses
to Notch/Delta signaling \cite{Bailey:1995rs, Ray:1996pf, Lecourtois:1997fi,
Morel:2003oe}, while Snail is a highly-conserved C$_2$H$_2$ zinc-finger TF that
represses activation in the mesoderm \cite{Gray:1994qm, Cowden:2002fk}. In
\mel{} and closely related species, NEEs also have a binding site for Dip-3
(Dorsal interacting protein-3) \cite{Erives:2004pd}, a Dorsal-binding protein
required for Dorsal/Twist synergistic activation and D/V patterning
\cite{Flores-Saaib:2000xw, Bhaskar:2000jl, Bhaskar:2002zt, Ratnaparkhi:2008ee}.
Besides these specialized binding sites, NEEs share distinct organizational
features pertaining to site placement, spacing, and polarity
\cite{Erives:2004pd}.  These observations suggest that NEEs form a distinct set
of sequences that ``read-out'' the Dorsal morphogen gradient at various
thresholds in the lateral regions of the embryo through specific protein
complexes composed of Dorsal, Twist, Snail, Su(H) and their co-factors.

Recently, we determined that the specialized NEE-type binding sites for Dorsal
and Twist have a unique function in setting the threshold for activation
\cite{Crocker:2008pb}.  In the NEEs from \mel{}, \pse{}, and \vir{}, we found
that: i) the precise length of a spacer DNA, which separates these well-defined
Dorsal and Twist binding sites, encodes the concentration threshold setting;
ii) natural selection has acted on the length of this spacer in different
lineages of the \Dros{} genus to adjust the threshold; and iii) these selective
\cis-regulatory adjustments have been performed at all NEEs across a given
genome, as would be expected if they are all co-evolving to a common change in
the \trans-morphogen gradient \cite{Crocker:2008pb}.  While this study
identified a heritable feature that encodes different responses to Dorsal, it
did not address its full functional range nor the function of the many other
Dorsal binding site variants, which constitute the clusters observed at these
enhancers. As such, it was not clear whether these additional Dorsal motifs
were necessary and/or sufficient for setting the gradient threshold,
participating in activation or repression, or any other regulatory function.

Here we test several wild-type and experimentally-modified NEEs from five
divergent species of \Dros: \mel, \ana, \pse, \wil, and \vir.  Importantly,
\ana{} and \wil{} represent the largest assembled \Dros{} genomes and are less
derived than the smaller, compact genomes of the melanogaster subgroup, which
may have lost important signatures indicative of past evolutionary history
\cite{Clark:2007ud}. Using this broad data set, we narrow the many explanations
of binding site clustering down to a single, unexpected, but ultimately
predictive hypothesis of concentration-threshold encoding, and explain several
perplexing constraints on the specialized sites of NEEs and their relative
organization. We show that complex enhancer clustering is a signature that ages
over time through a dynamic evolutionary process involving facile selection for
optimal threshold readouts and equally facile loss and/or selective deprecation
of former threshold-encodings. This process, which we term \term{dynamic
deprecation}, produces several non-functional signatures that obscure the
precise morphogen threshold-encoding mechanism that we functionally map and
confirm in this study. We conclude that the clustered signature observed in
most enhancers is produced by the dynamic evolutionary maintenance of the
accuracy of precise threshold-encodings.

\section*{Results}

\subsection*{Canonical NEEs are marked by \hcis-spectral clusters}

We found that binding site clusters at NEEs are characterized by a certain
``\cis-spectral'' signature, and refer to such clusters simply as
\term{\cis-spectra} (Fig.~1).  Binding site constituents of \cis-spectra are
revealed specifically within or immediately around the cluster as the motif
consensus for a TF is relaxed.  Thus, a \cis-spectral binding cluster remains
well-defined with increasing degeneracy of the binding motif.  For example, if
we use a \term{motif spectrum} of increasingly degenerate binding motifs
characteristic of Dorsal binding sites, we identify additional matching
sequences locally within the vicinity of the module, thus preserving the
definition of the cluster (Fig.~1, bottom rows of localized clustering). 

We defined three specialized \cis-element motifs that are associated with the
\cis-spectral clusters of canonical NEEs across \Dros{}: \SUH/\Da{}, \Db{}, and
\ECAT (see Fig.~1).  These motif signatures correspond to specialized versions
of more general binding motifs for Dorsal, Twist, Snail, and Su(H).
Importantly, the specialized motifs typically describe a single site at each
\cis-spectral cluster.

Despite the numerous binding site variants in Dorsal \cis-spectra, there are
only two distinct and separate, specialized Dorsal binding site motifs at each
NEE, here called \Da{} and \Db{}.  The specialized Dorsal binding motif \Da{}
partially overlaps an overly-determined and polarized Su(H) binding site \SUH{}
(Fig.~1).  In \mel, \SUH{} is polarized in the same direction as the $\mu$
site, a specialized binding site for Dip-3 \cite{Erives:2004pd}. Furthermore,
while the $\mu$ element appears to be absent in distant \Dros{} lineages,
\SUH{} is maintained in a polarized state, even after turnover events
\cite{Crocker:2008pb}. 

In contrast, the specialized Dorsal binding motif \Db{} is located uniquely
within $\sim$20~bp of the \ECAT{} element, the spacing to which encodes the
threshold response.  Furthermore, an invariant length-asymmetry in this nearly
palindromic \ECAT{} motif consistently points to \Db{} although \Db{} itself is
not polarized.  Importantly, we have never observed any Dorsal binding site
variant to be more tightly linked to the \ECAT{} element than the \Db{}
element.  The \ECAT{} element itself is a specialized \texttt{CA}-core E-box
(\fiveprime{CANNTG}) with an additional \texttt{T}, \ie{} the sequence
\fiveprime{CACATGT}. This \ECAT{} element is partially explained as the
superimposition of binding preferences for Twist and Snail.  Activating
Twist:Daughterless bHLH heterodimers bind the \DNA{YA}-core E-box
\fiveprime{CAYATG}, or \E{YA}, while the Snail repressor binds the motif
\fiveprime{SMMCWTGYBK} \cite{Castanon:2001kb, Gray:1994qm}. Thus, we predicted
that such a co-functional site may originate via selection for the superimposed
motifs, which corresponds to the sequence \fiveprime{SCACATGY}.  This
superimposed Twist/Snail binding motif is almost identical with the observed
\ECAT{} motif, \fiveprime{CACATGT}.  

We will refer to the three arranged elements of the polarized \ECAT{} site, the
spacer, and an unpolarized \Db{} site as an \EtoD{} \term{encoding}.  Using
this terminology, we will show that functional NEE modules need be composed
only of one \EtoD{} encoding, supported by a nearby generic Su(H) site. We will
show that the \EtoD{} sequence is the sole repository of the threshold encoding
variable at each NEE module, and that \cis-spectral clusters and certain
specialized sites are byproducts accumulated in mature enhancers. Last we will
show that an intrinsic mutational property of the \ECAT{} elements facilitates
the rapid selection of new \EtoD{} encodings.

\subsection*{Canonical NEEs from \hwil{} genome are enriched in \cis-spectra} 

To better understand the functional importance of multiple variant binding
sites for Dorsal and its co-factors within canonical NEEs, we analyzed the
\wil{} genome, which is the largest assembled \Dros{} genome (224 Mb)
\cite{Clark:2007ud}. The study of large genomes is important because relatively
compact genomes may have lost DNA signatures indicative of past evolutionary
processes.  The \wil{} lineage is an early branch of the same \SOPH{} subgenus
that includes the melanogaster subgroup, and represents $\sim$37 My of
evolutionary divergence since its common ancestor with \mel{}, whose genome has
been secondarily compacted (Fig.~2).  

To identify the canonical NEE set from \wil{}, it is sufficient to query the
genome for all 800~bp sequences containing the three motifs given by \SUH/\Da,
\Db, and \ECAT{}, without imposing any syntactical constraints, such as linked
Dorsal/Twist binding sites or polarized \SUH{} elements.  Such a query
identifies only the four canonical NEEs of \Dros{}, and these all conform to
the full syntactical rule set, despite significant levels of sequence
divergence.  We also verified that these NEE-bearing loci are expressed in the
neurogenic ectoderm of \wil{} embryos by whole-mount \insitu{} hybridization
(Fig.~3 A--D).  

We cloned DNAs encompassing the NEE sequences of \wil{} and tested them for
enhancer activity on a {\it lacZ} reporter stably integrated into multiple
independent lines of \mel{}.  Whole-mount \insitu{} hybridization of these
embryos with an anti-sense {\it lacZ} probe showed that the \wil{} enhancers
drive lateral ectodermal expression in \mel{} embryos (Fig.~3 E--H).  These
results demonstrate that these are functional enhancers present in loci
expressed in lateral regions of the neuroectoderm in \wil{} embryos.  In
general, \wil{} NEEs drive slightly narrower expression patterns in \mel{} than
their counterpart \mel{} NEE reporters, which may indicate that they are tuned
to higher threshold responses (Fig.~4).  

To determine whether the specialized Dorsal binding sites \Da{} and \Db{} are
embedded in clusters of Dorsal binding site variants as they are in other
lineages, we identified all sites in these sequences matching a Dorsal motif
spectrum  and found extremely dense Dorsal \cis-spectra in the NEEs of \wil{}
(Figs.~5--6). As quantified below, these are some of the densest clusters yet
seen in NEEs of the \Dros{} genus.  To ascertain whether the specialized Dorsal
motifs are maintained as unique copies in each NEE from \wil, or whether
additional Dorsal binding variant sites within each cluster also match these
specialized motifs as would be expected by random neutral drift
\cite{Kimura:1968ys,Ohta:1973uq}, we applied our path-finding method to
identify and characterize the most specialized Dorsal binding motifs within
their \cis-spectral clusters \cite{Erives:2004pd} (also see Supplement Part I).
We find that the \Da{} site occurs once in each NEE (Fig.~5).  Furthermore,
\Da{} continues to overlap the Su(H) binding site at this particular
specialized Dorsal binding site. This property is unique to \Da{} in canonical
NEEs across the \Dros{} genus.  Similarly, the \Db{} of \wil{} site occurs only
once in each NEE (Fig.~6). As expected, \Db{} is the closest variant Dorsal
binding site adjacent to \ECAT{} (Fig.~6).  The \Db{} consensus motif for the
canonical NEEs of \wil{} is nearly identical with the corresponding motif in
other previously-characterized lineages (Table~1).

The Dorsal \cis-spectral clusters of NEEs from \wil{} are associated with
another feature that is interesting in light of the reduced genomic deletion
rates relative to \mel: the \wil{} NEEs appear to be enriched in
\DNA{CA}-satellite sequence.  Given that the \ECAT{} sequence,
\fiveprime{CACATGT}, is composed entirely of \DNA{CA}-dinucleotide repeats, we
speculated whether the Dorsal \cis-spectra of NEEs are overlaid with a similar
\ECAT{} spectral cluster.  In support of this idea, we found several lengthy
\DNA{CA}-satellite tracts across the canonical NEE set of \wil{} (Fig.~7).
Almost all of these are associated with specific constituents of Dorsal
\cis-spectra.  Conversely, almost all constituent sites of Dorsal spectra are
associated with prominent \DNA{CA}-satellite tracts. For example, the
\cis-spectral cluster of the \NEE{vn} of \wil{} has an expanded
\DNA{CA}-satellite tracts associated with divergent \Db{} elements at
$\sim$340--400~bp and again at $\sim$580--630~bp, while the \wil{} \NEE{rho}
also has an expanded \DNA{CA}-satellite tracts coordinated to divergent \Db{}
elements at $\sim$130--150~bp and again at $\sim$270--290~bp (Fig.~7). Last,
the \NEE{vnd} sequence, which is the descendant of the oldest known NEE because
it is found in mosquitos, is characterized by the greatest number of lengthy
\DNA{CA}-satellite tracts in \wil{} (Fig.~7).

\subsection*{Constituents of \hcis-spectra represent non-functional necro-elements} 

In the \NEE{vnd} module of \wil{}, we detected the loss of one of two \EtoD{}
encodings that are present and intact in the \NEE{vnd} sequences from the \mel,
\pse, and \vir{} genomes \cite{Erives:2004pd,Crocker:2008pb}.  The first
\EtoD{} encoding has a tighter spacer compared to the second, distantly-spaced
\EtoD{} encoding.  Furthermore, the Dorsal binding site at this second encoding
is a divergent \Db{} element (Fig.~1).  In the \wil{} lineage, the \ECAT{}
element of this second divergent encoding expanded on both sides and then split
apart (Fig.~8A, inverted \DNA{CA}-satellite palindromic pair \#4).  This is
unambiguously an inactivating mutation of the Twist binding element.
Furthermore, the \NEE{vnd} of \wil{} is marked by several other such
palindromic tracts (numbered in Fig.~8A), of which the intact but also expanded
\ECAT{} site is the leftmost site in a series of increasingly-lengthy, split,
inverted palindromic \DNA{CA}-satellite repeats (Fig.~8B). These increasingly
expanded \DNA{CA}-satellite palindromes are associated with Dorsal binding site
variants that are increasingly divergent from the \Db{} consensus motif
(Fig.~8C).  

While the \wil{} \NEE{vnd} sequence has lost the second \ECAT{} site through
repeat expansion and separation of the two palindromic moieties, we did not
know whether this site functioned in species in which it is still intact. We
therefore tested two different fragments contained within a ``full-length''
949~bp enhancer sequence from the \vnd{} locus of \mel{} (Fig.~9A).  We tested
a 300~bp fragment that contains the first \EtoD{} encoding spaced by 10~bp, and
a 266~bp fragment that contains the second \EtoD{} encoding spaced by 20~bp.
Both fragments overlap and contain in common the extended \SUH/\Da{} site
(Fig.~9A).

We found that the 300~bp fragment works just as well as the 947~bp fragment
(Fig.~9 B, C, and E) while the 266~bp fragment hardly works at all (Fig.~9D and
4E).  Thus, the first \EtoD{} encoding, which is intact and tightly spaced, is
sufficient for the complete threshold-response, while the second \EtoD{}
encoding, which is expansively-spaced to a slightly divergent \Db{} element, is
non-functional.  We refer to the component sites of the second encoding as dead
elements, or \term{necro-elements}, and label them \NECAT{} and \NDb{}. While
the \NECAT{} sequence is intact, inspection of this \NDb{} sequence shows that
it has diverged somewhat from the genus-wide \Db{} consensus (Fig.~9F).  

These results indicated that Dorsal \cis-spectra and their associated
\DNA{CA}-satellite tracts are relic \EtoD{} encodings that were once functional
but eventually deprecated and replaced during lineage evolution.  While the
evolution of new encodings will sometimes occur via selection of spacer length
variants defined by existing elements, at other times it will occur via
selection of new replacement sites associated with new spacer lengths. Three
important features of \EtoD{} encodings increase the capacity for selection of
replacement encodings.  The first feature is the palindromic nature of the
\ECAT{} and \Db{} elements, which allows new \EtoD{} encodings to arise from
the selection of a single emergent site that is located on the other side of
its coordinating partner element in an existing encoding (a leapfrog).  The
second feature is that the \EtoD{} spacer range is broad-ranged and thus endows
functionality to sub-optimal encodings. The third feature is that
\DNA{CA}-dinucleotide satellite sequence is susceptible to repeat expansions
and contractions across the \Dros{} genus \cite{Schlotterer:2000dw,
Harr:2000fx, Harr:2000ud}.  We assume that the \ECAT{} sequence
\fiveprime{CACATGT} is dynamically unstable in NEEs because this element is
composed entirely of \DNA{CA}-repeats.  In support of this, we found that
intact \ECAT{} elements in the NEEs of several \Dros{} genomes are frequently
repeat-expanded beyond the core heptamer such that it matches the general
pattern given by \fiveprime(\DNA{CA})$_n$\DNA{T}(\DNA{GT})$_m$, where $n \ge 2$
and $m \ge 1$ (Table~2).  This is pronounced particularly in the larger,
uncompacted \ana{}, \wil{}, and \vir{} genomes, (Table~2).  These observations
are of utmost significance: spacer length variants produced by an intrinsic
repeat instability of the \ECAT{} element will drive different
threshold-responses.  This eventuality would also explain the highly invariant
nature of the \ECAT{} sequence.  Newly-selected replacement Twist/Snail binding
sites will evolve at target sequences most closely resembling the
dual-functioning site predicted by superimposed binding preferences (Fig.~10).
Initially, such an emergent site will be associated with a suboptimal spacer.
However, random neutral drift to the specific \ECAT{} sequence would result in
the availability of spacer length variants via \DNA{CA}-satellite repeat
expansion/contraction. Thereafter, frequent occasions for selection of spacer
variants produced by such a site would result in the apparent ``constraint'' of
the Twist element. 

The evolution of threshold readouts via dynamic deprecation and replacement of
encodings, as facilitated by instrinsic \ECAT{} instability, makes several
testable predictions.  First, a dynamic deprecation model is supported if
longer \DNA{CA}-satellite tracts in \wil{} NEEs are loosely associated with
specific components of Dorsal \cis-spectra, especially when they are spaced
beyond the functional range of the spacer element.  Second, necro-element
accumulation may progress in a clock-like fashion followed by neutral
divergence of these sites.  Thus \cis-element spectra for both Dorsal and Twist
binding motifs should be associated with mature NEEs that are canonical to the
lineage, but not in newer NEEs that might have arisen more recently.  Third, we
should find that threshold readout is correlated to spacer length but not to binding
site density. Fourth, we should be able to remove deprecated encodings without
affecting the threshold readout (as in Fig.~9 C and E).  Conversely isolated
deprecated encodings should not possess lower thresholds compared to the intact
enhancer (as in Fig.~9 D and E).  

\subsection*{Canonical NEEs across \hDros{} are enriched in necro-element spectra} 

To address the generality of \DNA{CA}-satellite accumulation in NEEs across the genus,
we checked the percentage of \DNA{CA}-satellite in NEEs from 
\mel{}, \pse{}, \wil{}, and \vir{} relative to their genomic
background levels (Table~3).  These analyses consistently show that
\DNA{CA}-satellite is enriched in NEEs above genomic background rates.
Importantly, this elevated level is not due to the presence of intact \ECAT{}
motifs, which constitute only a minor fraction of \DNA{CA}-repeat sequence in
NEEs (Table~3).

To address the possibility that elevated \DNA{CA}-satellite composition is a
feature common to developmental enhancers, we then looked at several canonical
enhancers that respond to the Bicoid morphogen gradient, which patterns the
anterior/posterior (A/P) axis. We identified the {\it hunchback} ({\it hb})
enhancers, the {\it giant} ({\it gt}) posterior enhancers, the {\it Kruppel}
({\it Kr}) enhancers, and the well-studied {\it even-skipped} ({\it eve})
stripe 2 enhancers from each of 4 genomes: \mel, \pse, \wil, and \vir. All of
these enhancers are active in the same embryonic nuclei as the NEEs and thus
constitute a well-matched control group. We found that while all 16 of these
A/P enhancers possess evolving clusters of Bicoid binding site spectra (data
not shown), none of them possess the elevated \DNA{CA}-satellite levels that
characterize canonical NEEs from these same species (Fig.~11). Thus, there is a
tremendous sequence bias that is unique to canonical NEEs across the genus and
in stark contrast to the sequence composition of both their genomes and other
non-NEE enhancer clusters. Furthermore, this NEE compositional bias is related
to specific functional elements employed by NEEs.

Having found we could identify the extent of Dorsal \cis-spectra with
confidence, we then checked its potential to encode or influence Dorsal
concentration threshold read-out of NEEs. For example, we checked the relation
between threshold-readout and the density of Dorsal halfsites in a region
anchored $\pm$480~bp from \Db{} (Fig.~12A).  For this we measured the stripe
width at 50$\%$ egg length as measured by the number of nuclei expressing the
reporter gene from the ventral border of expression up to the dorsal border.
We also found no relation between Dorsal binding site densities and
threshold-encodings after trying diverse other descriptors of a Dorsal binding
site (data not shown).  Identical densities of Dorsal halfsites, degenerate
full-sites, and more complete full-sites are present in different enhancers
that read-out different Dorsal concentration thresholds and vice versa.

In contrast, if we plot the length of the \EtoD{} spacers for NEEs with
unambiguous \EtoD{} encodings (\ie, encodings with single intact \ECAT{} and
\Db{} elements) and except those from the dorsally-repressed \vnd{} loci, we
see a well-defined, hump-shaped curve, whose peak activity tops at around 7~bp
and falls on either side of this maximum. The spacer elements from the
consistently high-threshold \NEE{vnd} sequences across the genus obey a
similar, albeit depressed, curve because of one additional regulatory input
(data not shown).  Thus, the elevated \DNA{CA}-satellite content and its
associated Dorsal \cis-spectra are consistent with the central hypothesis that
the sequence composition of these enhancers has been shaped by a long history
of repeated deprecation and compensatory selection of \EtoD{} encodings by a
process which has been active for more than 200 My in the case of the \NEE{vnd}
sequence, and more than 40 My at other canonical NEEs. 

Given the extent of \cis-spectral signatures associated with Dorsal and Twist
binding elements in mature NEEs, we asked whether the specialized \Da{}
site, which overlaps an unusually specialized Su(H) binding site, might also be
a \Db{} necro-element that was conveniently turned into a Su(H) site.  To
address this question, we first compared the \Da{} and \Db{} consensi motifs
across all five divergent \Dros{} lineages for which we functionally tested
NEEs in \mel{} (Table~1, Fig.~13A).  Remarkably, we find that the second half
of the \Da{} has diverged across the genus faster than the first half. This
second half is the portion that does not overlap the Su(H) binding site.
Unlike, the slight lineage-specific variations of \Db{}, \Da{} motif divergence
can be characterized as increasingly degenerate when departing from the
ancestral \Da{} motif, which is closest to a \Db{} motif itself.  

To test whether the Su(H) binding site is itself functional and perhaps the
principal reason for persistence of the ``ghost'' \Da{} motif, we specifically
mutated the Su(H)-specific portion of the \SUH/\Da{} site in the \NEE{rho} of
\mel{} (Fig.~13 A and C).  This specific mutation appears to weaken the
activation response of the enhancer without affecting the specific threshold
setting (Fig.~13 B--C).  Because we have shown a general tendency of functional
\ECAT{} elements to have expanded beyond the heptamer sequence (Table~2), and
of deprecated \ECAT{} elements to have experienced runaway expansion into
longer tracts (Figs.~7--8), we suspect that this process tends to push away
combinatorial enhancer elements, such as Su(H) binding sites.  In this context,
selection may favor new Su(H) binding sites that are closer to the current
functional encoding.  Conveniently, deprecated \NDb{} sequences are similar to
sequences matching the Su(H) binding motif and thus provide a convenient set of
target sites for re-evolving more proximal Su(H) sites.

\subsection*{Newly evolved NEEs are not enriched in \cis-spectra} 

Our results on the canonical NEEs of the four divergent lineages of \mel, \pse,
\wil, and \vir{} NEEs demonstrate that much of their sequence composition
corresponds to relic deprecated encodings. This pertains not only to the
sequences in between intact Dorsal, Twist/Snail, and Su(H) binding motifs but
to most of the recognizable and intact TF sites and variants as well. Because
we predict that necro-element accumulation is a neutral signature related to
the number of past threshold adaptations, whose number likely increases with
age, we were curious about the extent of \cis-spectral signatures in younger
NEEs.  We previously documented a new NEE sequence at the \sog{} locus of
\mel{} \cite{Crocker:2008pb}.  The \mel{} \NEE{sog} sequence has a
\DNA{CA}-dinucleotide content of 14.4$\%$, which is on par with highest levels
seen in A/P enhancers from all lineages but is mid-range for NEEs from \mel{}
(compare with Fig.~11B points in the A/P box). However, because the
\DNA{CA}-content of NEEs from \mel{} may have been secondarily reduced, we
therefore wanted to query uncompacted \Dros{} genomes with a parameter set that
is constrained only by the minimal molecular requirements.  Thus, we queried
the two largest \Dros{} genome assemblies, which corresponded to \ana{} (231.0
Mb) and \wil{} (235.5 Mb). Both of these species are in the \SOPH{} subgenus,
which includes \mel{}.

Of the 1~kb genomic windows centered on all \Db{} instances in any given genome
and containing \ECAT{} anywhere in that window, we identified the subset of
these sequences that also contained a generic (``un-specialized'') Su(H)
binding site as well as linked Dorsal and Twist binding elements.  The generic
Su(H) site replaces the composite extended motif that described an
overly-determined \SUH{} element and the overlapping \Da{} ghost site.  Using
this set of minimal criteria, we nonetheless were able to identify the
canonical NEE repertoires for each species.  

From the \ana{} genome, we identified, cloned and tested both a functional set
of canonical NEEs (Fig.~14), and a new NEE at the \Dl{} (Dl) locus (Fig.~15).
Delta encodes a ligand for the Notch receptor, whose signaling is relayed by
the Su(H) TF itself \cite{Lecourtois:1997fi, Lecourtois:1998xd}.  In \mel{}
embryos, Delta is expressed in a narrow lateral stripe in the mesectoderm and
ventral most row of the neurogenic ectoderm using sequences that are unrelated
to the \NEE{Delta} sequence of \ana{} \cite{Morel:2003oe}.  

Like the \NEE{sog} sequence, which matured in the melanogaster subgroup, the
\NEE{Delta} sequence in \ana{} has not yet accumulated either
\DNA{CA}-satellite content or the Dorsal \cis-spectra characteristic of
necro-element plaques (Fig.~15A).  Nonetheless, this enhancer is functional in
\mel{} embryos (Fig.~15B).  Inspection of its Su(H) binding site reveals that it
does not overlap a ghost \Da{} motif, which demonstrates again that \Da{} is
not required (Fig.~15C).  This is consistent with the interpretation that \Da{}
motifs are deprecated \Db{} motifs exapted into functional \SUH{} elements at
mature NEEs, whose sequence compositions have been biased by long histories of
necro-element accumulation.  

The \NEE{Delta} enhancer has a spacer of 3~bp, and occupies the low-end of the
threshold mapping function (Fig.~12). Therefore, because we characterized both
high and low threshold NEEs that have evolved more recently in the \Dl{} and
\sog{} loci of \ana{} and \mel{}, respectively, without much necro-element
accumulation, the \cis-spectra of mature NEEs are likely unrelated to function.
Instead, the absence of the necro-element plaques suggest a shorter period of
evolutionary maintenance, consistent with their phylogenetic distribution.

\section*{Discussion} 

In this study of regulatory DNAs from the \Dros{} genus, we found that a
certain Dorsal-threshold encoding mechanism maps a spacer length of 3--15~bp,
which links a pair of well-defined Dorsal and Twist binding sites, onto one
well-defined dorsal border of expression that is 5--15 nuclei past the ventral
border of the neurogenic ectoderm. The specialized Twist-binding \ECAT{}
sequence is a constrained motif that satisfies binding preferences for both the
Twist activator and the Snail mesodermal repressor.  This sequence is also a
palindromic \DNA{CA}-satellite sequence that is prone to \DNA{CA}-dinucleotide
repeat expansions that alter the precise threshold setting spacer.  Natural
selection acts continuously to exploit \ECAT{} instability to adapt the
precise, threshold-setting spacers between adjacent and intact Dorsal and Twist
binding elements.  This process may also accelerate site turnover, because it
would frequently necessitate stabilizing selection of compensatory threshold
settings in response to this intrinsic instability.  Thus, evolutionary
maintenance of optimal NEE function involves the clock-like production of dead
Dorsal and Twist binding elements, which we call necro-elements.  Necro-element
accumulation is the major determinant of sequence composition in enhancers that
have matured beyond a certain age ($>$10 My).  Further genomic sampling of taxa
will allow refinement of the necro-element clock, and ascertain whether it
reaches a saturation point for the most ancient enhancers. This question
increases the need to sequence larger genomes that are not compressed
secondarily by high deletion rates \cite{Peterson:2009ph}.  

We found that the specialized Su(H) binding site \SUH{} is exapted from
deprecated, non-functional Dorsal binding sites in all canonical NEEs of \Dros.
\SUH{} appears to influence the strength of activation without affecting the
Dorsal concentration-threshold response. This site is specialized in mature
NEEs but not in more recently evolved NEEs. This unusual turnover process for
Su(H) sites may be necessitated by the tendency of \DNA{CA}-satellite expansion
to act as a ``conveyor belt'' pushing out coordinating elements such as the
Su(H) binding site, but leaving a convenient path of deprecated elements that
are easily exapted into closer Su(H) sites.  

We found that functional NEEs can be derived from truncated fragments of
mature NEEs that lack necro-elements while continuing to encode the correct
threshold setting.  Also, functional NEEs have evolved more recently at
non-canonical loci without having yet accumulated the characteristic
necro-element plaques seen in older NEEs.  Such NEEs bear Su(H) sites that do
not extend to deprecated, ghost Dorsal binding sites.  

Last, we found a smooth continuum between intact NEE elements and increasingly
divergent deprecated necro-elements in these enhancers. Furthermore, because
the extreme range of this continuum is associated with the age of the enhancer,
we infer that necro-element accumulation begins with each NEE origination and
is continuously co-extant with its adaptive maintenance. This has led us to a
richly-predictive yet parsimonious model of NEE evolution that we call
\term{dynamic deprecation} (Fig.~16).  With increasing time, the background
sequence composition of enhancers is profoundly altered and eventually
dominates the nature of binding site sequences because it provides a
highly-biased ground state from which new sites are exapted.  

\paragraph{Defining necro-elements, \hcis-spectra, and deprecated
necro-elements.} We have used the term {\it necro-element} initially to
describe intact or nearly intact binding sites occurring within well-defined
clusters but which are no longer relevant in the current threshold encoding.
This term can be applied to sites subjected to dynamic deprecation, including
those that are deprecated solely through changes in syntax.  However, because
there is no clear dividing line between potentially-functional binding sites
deprecated by syntax and increasingly divergent sites, we have chosen to expand
the use of ``necro-element'' to refer to the entire continuum constituting a
clustered plaque of necro-elements.  We call such clusters {\it \cis-spectra}
in order to distinguish them from functional ``clusters'' of binding sites.
{\it Cis}-spectra are well-defined operationally as motif clusters that remain
distinct from background genomic sequence as the degeneracy of the matching
motif is increased and additional, presumably older, relic sites are revealed.
In this context, we used the term \term{motif spectrum} to refer to the
bioinformatic set of motif descriptors that detect \cis-spectra for a given TF. 

The use of the prefix \term{necro-} rather than the prefix \term{pseudo-} is
justified by several important distinctions that are peculiar to
necro-elements.  Etymologically, the Greek root
\begin{otherlanguage*}{greek}yeudo-\end{otherlanguage*} means `false', while
the Greek root \begin{otherlanguage*}{greek}nekro-\end{otherlanguage*} means
`dead' and more accurately connotes `loss of function'. This is an important
distinction because biological systems are rich in functional dissimulation
(\eg, mimicry and camouflage on an organismal scale, but also extending to
viral oncogenes that dissimulate normal cellular genes, and potentially true
pseudo-elements that function as decoy DNA elements to sequester a
transcription factor).  Biologically, the chosen term must encompass in its
definition both deprecated and non-deprecated elements, as well as both
non-functional and functionally-redundant elements.  Conventionally, the usage
of the \term{pseudo-} prefix for sequence lengths on the length-scale of
\cis-elements is unwieldy because it is used almost exclusively for
recognizable homologs of protein-coding genes with clear inactivating mutations
(\eg, internal stop codons, and frameshifts).  Necro-elements cannot always be
identified by sequence alone because they can be rendered functionally
redundant or non-functional by selection on syntax. 

We also used the term \term{deprecation} to connote additional information as
to the probable role of selection in producing a necro-element.  A
\term{deprecated necro-element} is a useful distinction to characterize a
necro-element that has undergone selection for attenuated or complete loss of
function in connection with the selection of a replacement threshold-encoding
located either at the enhancer or elsewhere in the locus. Thus, deprecation
implies that selection was active in removing an epistatic relationship between
two conflicting threshold-encodings. Selection may favor such an outcome when
the pre-deprecated functional element encodes a lower-threshold than the
positively selected replacement encoding. In such cases, a low threshold
encoding masks the function of any high threshold encoding under positive
selection and must engender active selective deprecation. On the other hand, if
a high-threshold encoding is being selectively replaced by a low threshold
encoding, we expect no active deprecation forces. Instead, we expect gradual
loss of function via neutral drift. This is an unexpectedly novel evolutionary
mechanism for generating apparent regulatory redundancy. In this context, we
suggest that redundant ``shadow enhancers'', which have been observed at
several Dorsal target loci in the \mel{} genome \cite{Hong:2008ly}, should be
incorporated into the same dynamic deprecation framework when appropriate.
Selection may adapt an existing threshold encoding or transition its focus to a
new threshold encoding that is located either within the same enhancer or
elsewhere in the locus.  Multiple such events are likely to pepper the
idiosyncratic histories of different lineages at different times.  In this
context, \term{shadow enhancers} may be defined as out-moded enhancers, which
were either redundant when replaced by distant low threshold enhancers, or
actively deprecated by selection until their threshold was at least higher than
a newer optimal low threshold enhancer located elsewhere in the same locus. 

\paragraph{Summary and implications.} In principle, \cis-spectral plaques of
necro-elements should accumulate in all complex eukaryotic enhancers that
encode key regulatory variables in a precise syntax. The extent of this
clustering would then be determined by the age of the enhancer, and the number
or rate of replacement adaptations over this time. While many of the intensely
studied enhancers of \Dros{} have corresponded to early embryonic enhancers that
are evolutionarily sensitive to changes in egg size and morphology,
they are also proving useful in untangling the molecular and evolutionary
aspects of enhancer biology. 

In this evolutionary context, the biology of necro-element spectra of D/V
enhancers appears to be directly applicable to A/P enhancers responsive to the
Bicoid morphogen gradient system.  Evolution of egg size and developmental
timing during embryogenesis is likely to place evolutionary demands on both A/P
and D/V morphogen gradient systems, which are operating simultaneously in the
same cells.  While we have shown that Dorsal binding site density does not
correlate with threshold encoding, others have shown that Bicoid binding site
strength in the heavily-clustered A/P enhancers does not correlate with A/P
position of activity \cite{Ochoa-Espinosa:2005vl}.  Under the dynamic
deprecation theory of enhancer evolution, this paradox is explained if the
majority of Bicoid binding site variants at such clusters represent
necro-elements deprecated by mutations affecting the site itself, its
coordinating site(s), and/or their syntactical relation. This interpretation
can be confirmed by future studies identifying the minimal molecular requirements
for encoding variable Bicoid-response thresholds.

One important implication for current studies is that motif descriptors and
algorithmic motif predictors should be constructed over a judiciously-chosen
set of functionally-equivalent sites across a genome, rather than on the
continuum of necro-element spectra at a cluster.  Such clusters are often
exploited statistically to increase the number of ``example elements''.  Such
approaches lead to degenerate motifs describing both extant functional elements
and surrounding deprecated sequences.  Newer approaches that are both
alignment-free and wary of exploiting the abundance of related sequences will
do better at distinguishing functional elements from evolutionary artifact
\cite{Erives:2004pd, Leung:2009sf}.

The conceptual re-framing of the functional evolution of enhancers overturns a
common assumption that all binding site variations within an enhancer are
functional and/or subtly necessary. This assumption has been directly
responsible for the impression that the ``cluster code'' is ``flexible'', by
which is meant that enhancer activity is robust to mutational disruption
\cite{Brown:2007uk, Hare:2008fu, Liberman:2009rp, Cameron:2009mz}.  However,
whether these site sequences are flexible or not flexible is only a productive
question if the observed sequences are functional in some way.  In contrast,
our results have supported the existence of a precise encoding scheme that uses
only a {\it limited subset} of sites in the cluster \cite{Erives:2004pd}.
Mutational variation in the organization of these specialized sites produces a
specific and well-mapped range of expression phenotypes \cite{Crocker:2008pb}.
Indeed, because this precise encoding scheme turns brittle when extended past
its functional range, selective deprecation is facilitated.  This view is
further enriched by considering the complex macroevolutionary processes that
result when taxa and lineages persist through several expansions caused by
non-static ecological/climatic conditions
\cite{Hewitt:2003rt,Hewitt:2004fk,Hewitt:2004uq,McPeek:2008an}. Regulatory
evolution is likely to underlie many of the stabilizing and adaptive changes
associated not only with these climate-driven historical events but future
climate changes as well \cite{Overpeck:1991kx}. 

The potential for gene regulatory evolution is likelier when encoding schemes
for relevant regulatory traits are broad-ranged functions that map genotype
(enhancer sequence) to phenotype (expression profile). Precise codes provide
the additional category of syntax on which natural selection can act. However,
a broad or evolutionarily-varied phenotypic range may be a simple consequence
of molecular mechanisms that are employed ontogenetically at multiple loci in
precise but varied functional configurations. Understanding this complex
relation between molecular encoding systems and their complex evolutionary
histories may prove useful in gauging the intrinsic adaptive potential of
specific systems subjected to future climate change \cite{Klausmeyer:2009uq}.

% You may title this section "Methods" or "Models".  "Models" is not a valid
% title for PLoS ONE authors. However, PLoS ONE authors may use "Analysis" 

\section*{Materials and Methods}

\paragraph{Embryonic experiments.} Animal rearing, P-element mediated
transformations, embryonic collections, staging, anti-DigU probe synthesis, and
whole-mount \insitu{} hybridizations were conducted as previously reported
\cite{Crocker:2008pb}.

\paragraph{Probes for whole-mount \insitu{} hybridization in \hwil{} embryos.}
Primers for probe synthesis are as listed here. \rhomb: \fiveprime{CCGCC TTTGC
CTATG ACCGT TATAC AATGC} and \fiveprime{Pr-TTAGG ACACA CCCAA GTCGT GC}, where
\DNA{Pr} = the T7 promoter sequence \fiveprime{CCGCC TAATA CGACT CACTA TAGGG}.
\vn: \fiveprime{CCGCC TAGTG ACGAC AACAA CAACA GTAGC} and \fiveprime{Pr-ATTTT
CACTCA CAGCC ATTTT CACC}.  \vnd: \fiveprime{CCGCC CTAGT CCGGA TAGCA CTTCG C}
and \fiveprime{Pr-CGGCT GCCAC ATGTT GATAG G}.  \brk: \fiveprime{CCGCC AACAA
AGTTC GTCGG CAACA ACG} and \fiveprime{Pr-CATGG TGAGG TGAGG ACTAT GG}.

\paragraph{Whole genome sequence analysis.} Current versions for all genomes
were downloaded from Flybase (www.flybase.org) and these correspond to assembly versions:
dmel ver5.22, dana ver1.3, dpse. ver2.6, dwil ver1.3, and dvir ver1.2.  Various
whole-genome queries were conducted using shell scripts composed of shell,
perl, grep, and wc UNIX commands and are available upon request. Separate
queries were conducted for NEE signatures and \DNA{CA}-satellite content.
Special genome files were processed for counting percent content of a given
motif.  We call these ``$\ast$.HNF'' files because they are header and N-free
files; these having been replaced by newline characters.

% Do NOT remove this, even if you are not including acknowledgments
\section*{Acknowledgments}

The authors thank Michael Dietrich, Mark McPeek, Alysha Heimberg, Kevin
Peterson, Lisa Fleischer, Ilya Ruvinsky, and Bryan Kolaczkowski for reading and
commenting on earlier versions of the manuscript. The authors also thank Ann
Lavanway for technical assistance. Completion of this work was supported by an
NSF CAREER award to A.E. (IOS 0952743).

\ifthenelse{\boolean{publ}}{\end{multicols}}{}

%\section*{References}
\bibliography{erives}
\clearpage

\section*{Figures}

\begin{figure}[!ht]
\begin{center}
\includegraphics[width=5.5in]{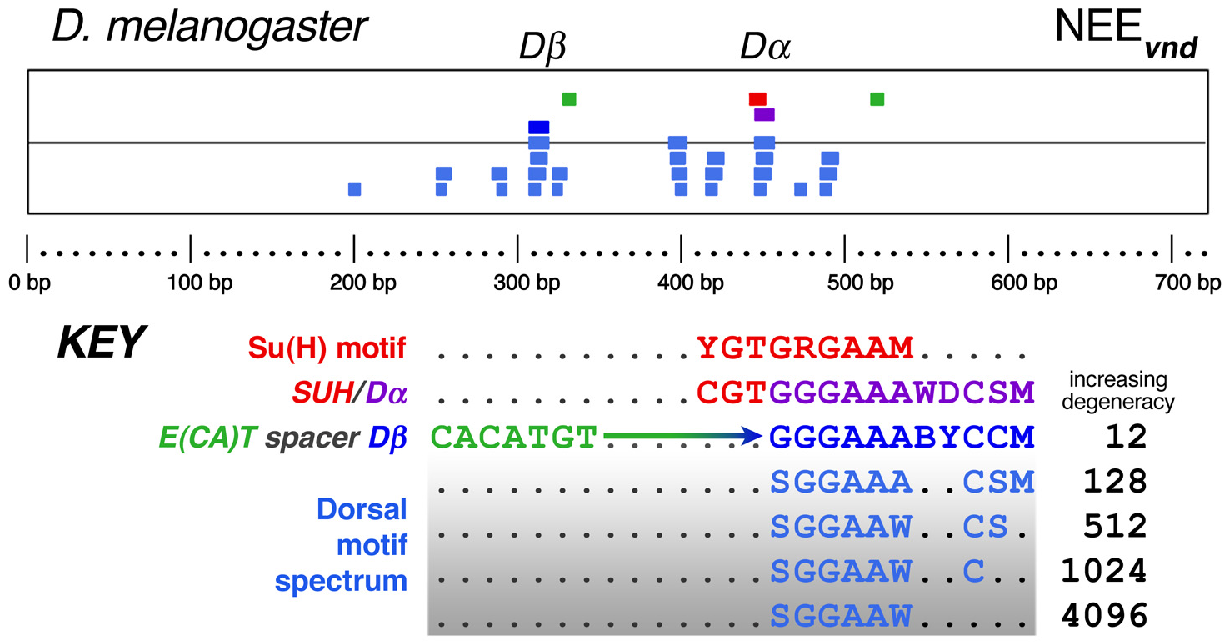}
\end{center}
\caption{
{\bf Organization of specialized elements within Dorsal \textit{cis}-spectra of canonical NEEs.}\\ 
Shown are the specialized sites embedded within the Dorsal \cis-spectra of the
\mel{} \NEE{vnd} sequence, which is representative of canonical NEEs at the
\rhomb, \vn, \vnd, and \brk{} loci of the \Dros{} genus.  Numerous lines of
evidence in this study demonstrate that the Dorsal \cis-spectra are specific to
mature NEEs ($>$40 My old), non-functional, and likely produced by dynamic
deprecation of precisely spaced Dorsal and Twist sites.  Dorsal \cis-spectra
are defined by a motif spectrum of increasingly degenerate Dorsal binding
motifs.  All instances of the motifs listed in the key are shown in the
graphic.  The motif sequences in all of the figures and text are written
according to IUPAC DNA convention: \DNA{S} = [\DNA{CG}], \DNA{W} = [\DNA{AT}],
\DNA{R} = [\DNA{AG}], \DNA{Y} = [\DNA{CT}], \DNA{K} = [\DNA{GT}], \DNA{M} =
[\DNA{CT}], \DNA{B} = [\DNA{CGT}], \DNA{D} = [\DNA{AGT}], \DNA{H} =
[\DNA{ACT}], \DNA{V} = [\DNA{ACT}], \DNA{N} = [\DNA{ACGT}], where nucleotides
in brackets are equivalent.  All Dorsal binding sites, motifs, and variants
will be depicted with the best halfsite on the \fiveprime{} side regardless of its
polarity to \ECAT.
}
\label{fig1}
\end{figure}\clearpage

\begin{figure}[!ht]
\begin{center}
\includegraphics[width=5.5in]{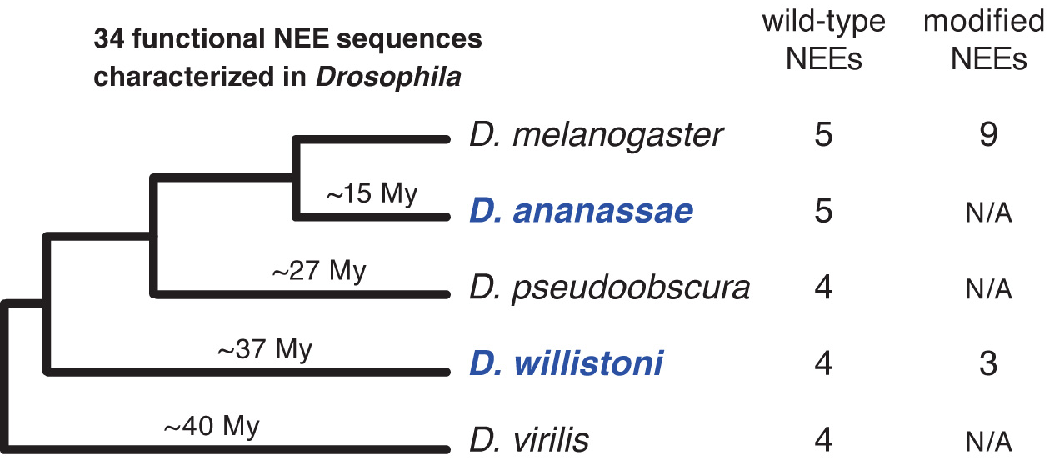}
\end{center}
\caption{
{\bf \hDros{} phylogeny with tested NEE sequences.}\\
In this study, we expand our previous studies to two genomes not marked by secondarily-derived
compact genome sizes. These genomes correspond to the \ana{} and \wil{} lineages (blue). We also 
expand our analyses by testing additional mutated versions of these and previously cloned enhancers.
}
\label{fig2}
\end{figure}\clearpage

\begin{figure}[!ht]
\begin{center}
\includegraphics[width=5.5in]{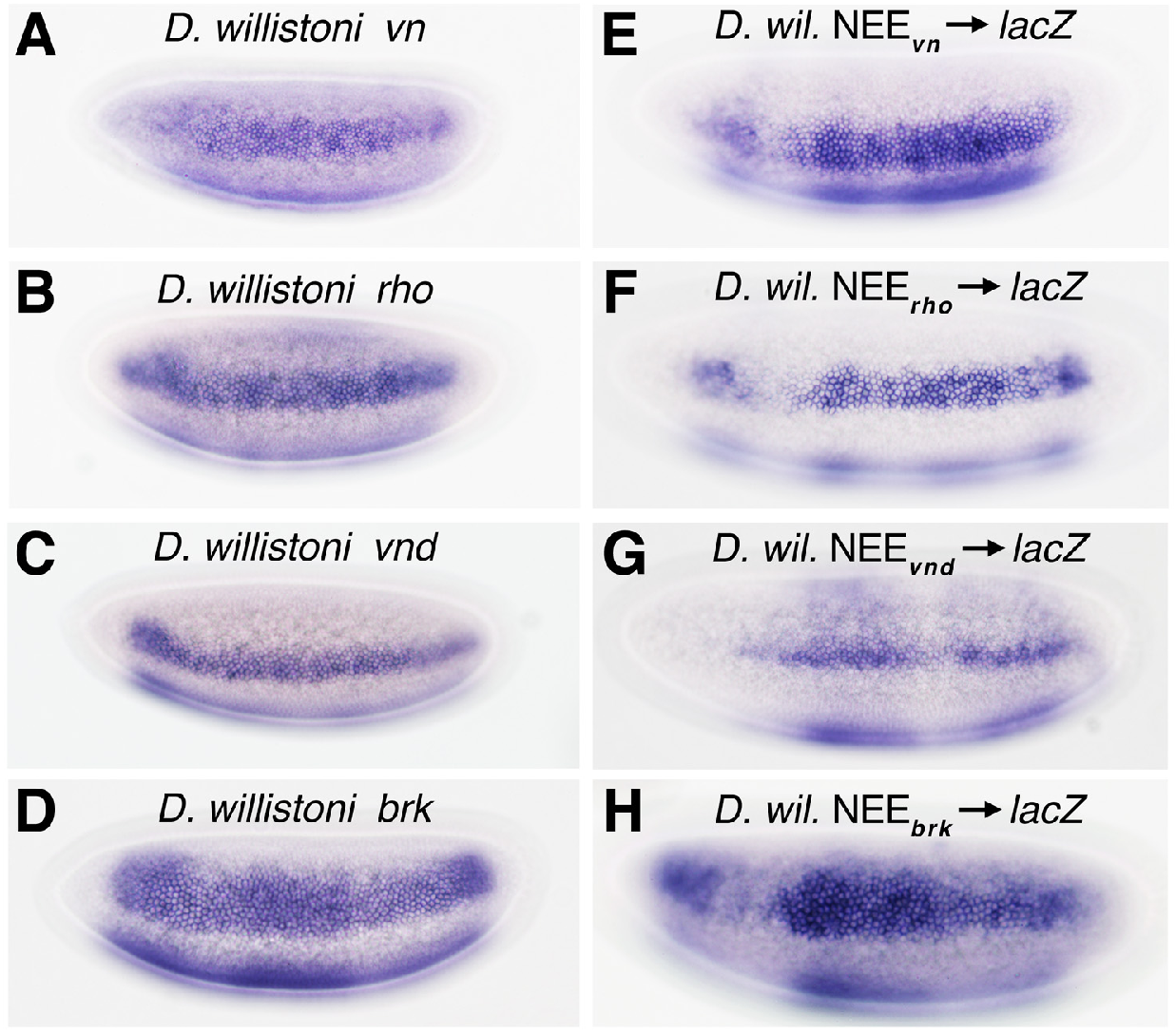}
\end{center}
\caption{
{\bf Functional NEEs from \hwil.}\\
Functional NEEs from \wil{} occur in canonical loci that are also expressed in
the neurogenic ectoderm.  {\bf A--D)} NEE-bearing loci in \wil{} are expressed
endogenously in the neuroectoderm of stage 5(2) embryos as shown by \insitu{}
hybridization with an anti-sense RNA probe to exonic sequences.  {\bf E--H)}
NEE sequences from \wil{} can drive a {\it lacZ} reporter gene in transgenic
\mel{} embryos as shown by \insitu{} hybridization with an anti-sense RNA probe
to {\it lacZ}.  Embryos in all figures are depicted with anterior pole to the
left, and dorsal side on top. Image labels indicate the species of the
embryo, and the gene or reporter being detected. All reporters are in \mel{} embryos.
}
\label{fig3}
\end{figure}\clearpage

\begin{figure}[!ht]
\begin{center}
\includegraphics[width=5.5in]{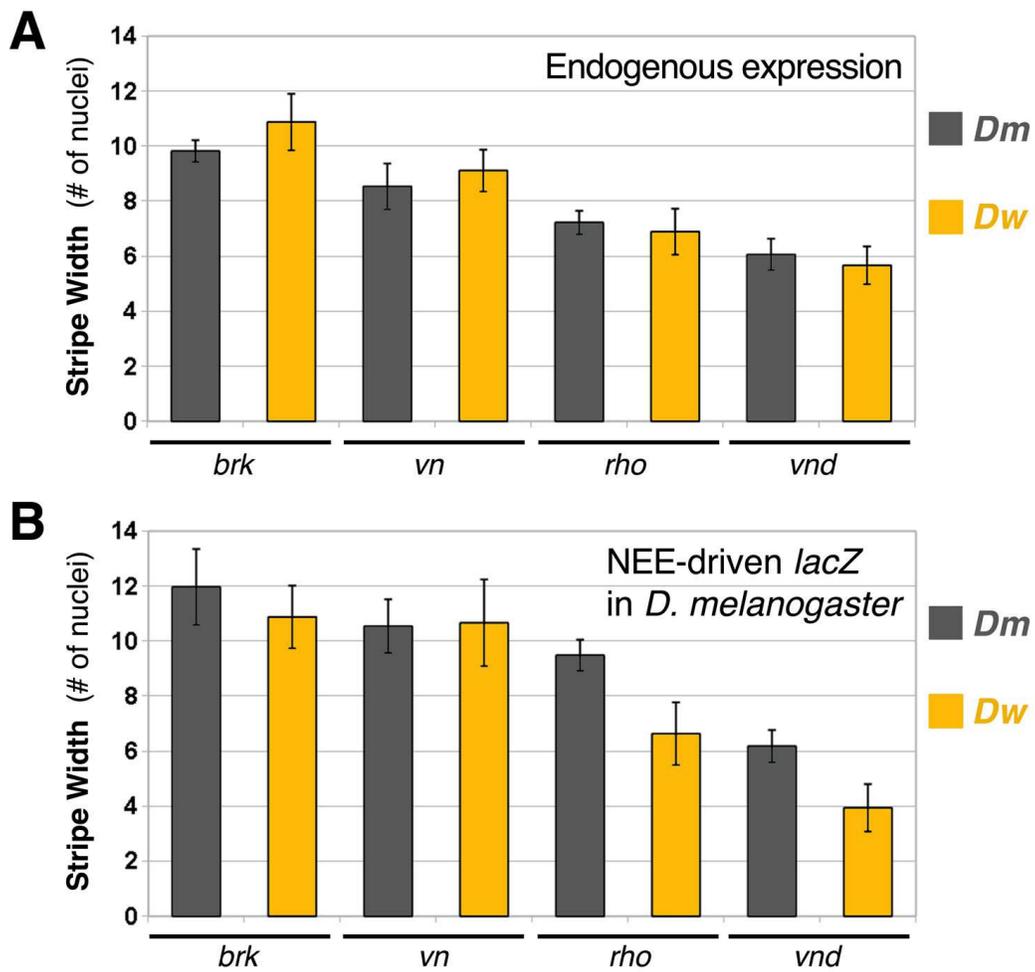}
\end{center}
\caption{
{\bf \hwil{} NEEs are set to higher concentration thresholds than \hmel{}.}\\
See text.
}
\label{fig4}
\end{figure}\clearpage

\begin{figure}[!ht]
\begin{center}
\includegraphics[width=5.5in]{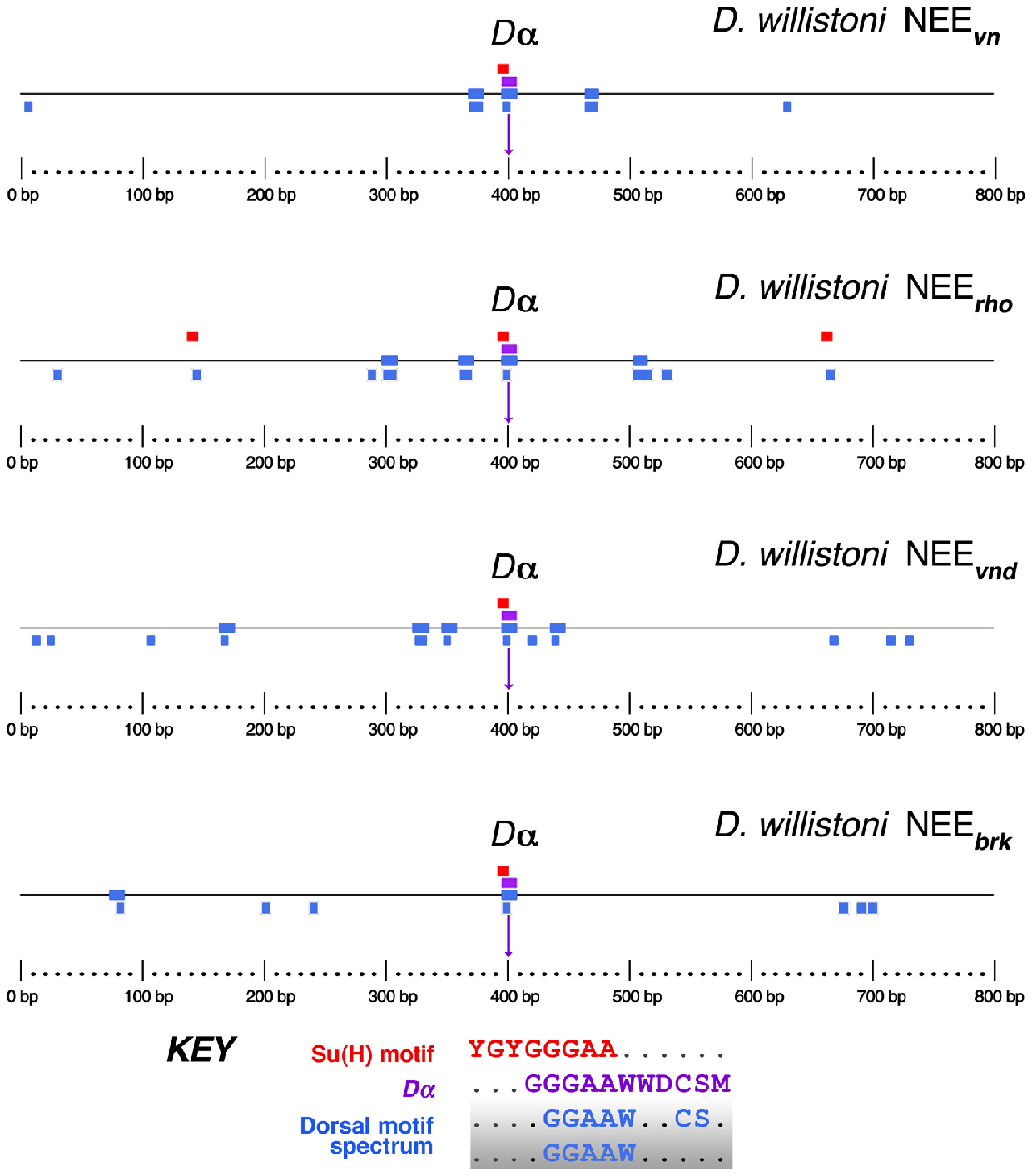}
\end{center}
\caption{
{\bf Dorsal \hcis-spectra in canonical \hwil{} NEEs contain a single \hDa{} site.}\\
Constituents of Dorsal \cis-spectra in \wil{} NEEs are visualized by matches to
Dorsal halfsites (base {\it D} halfsite, pale blue) and degenerate full sites
(base {\it D}, light blue) as shown in the key.  One such site at each
canonical NEE matches the \Da{} consensus (purple).  This same site overlaps a
Su(H) binding site (\SUH, red), which occurs on the top strand at each NEE. For
efficient referencing across the set, all NEEs from \wil{} are aligned and
centered on the unique \Da{} site, plus or minus 400~bp, unless otherwise
stated.
}
\label{fig5}
\end{figure}\clearpage

\begin{figure}[!ht]
\begin{center}
\includegraphics[width=5.5in]{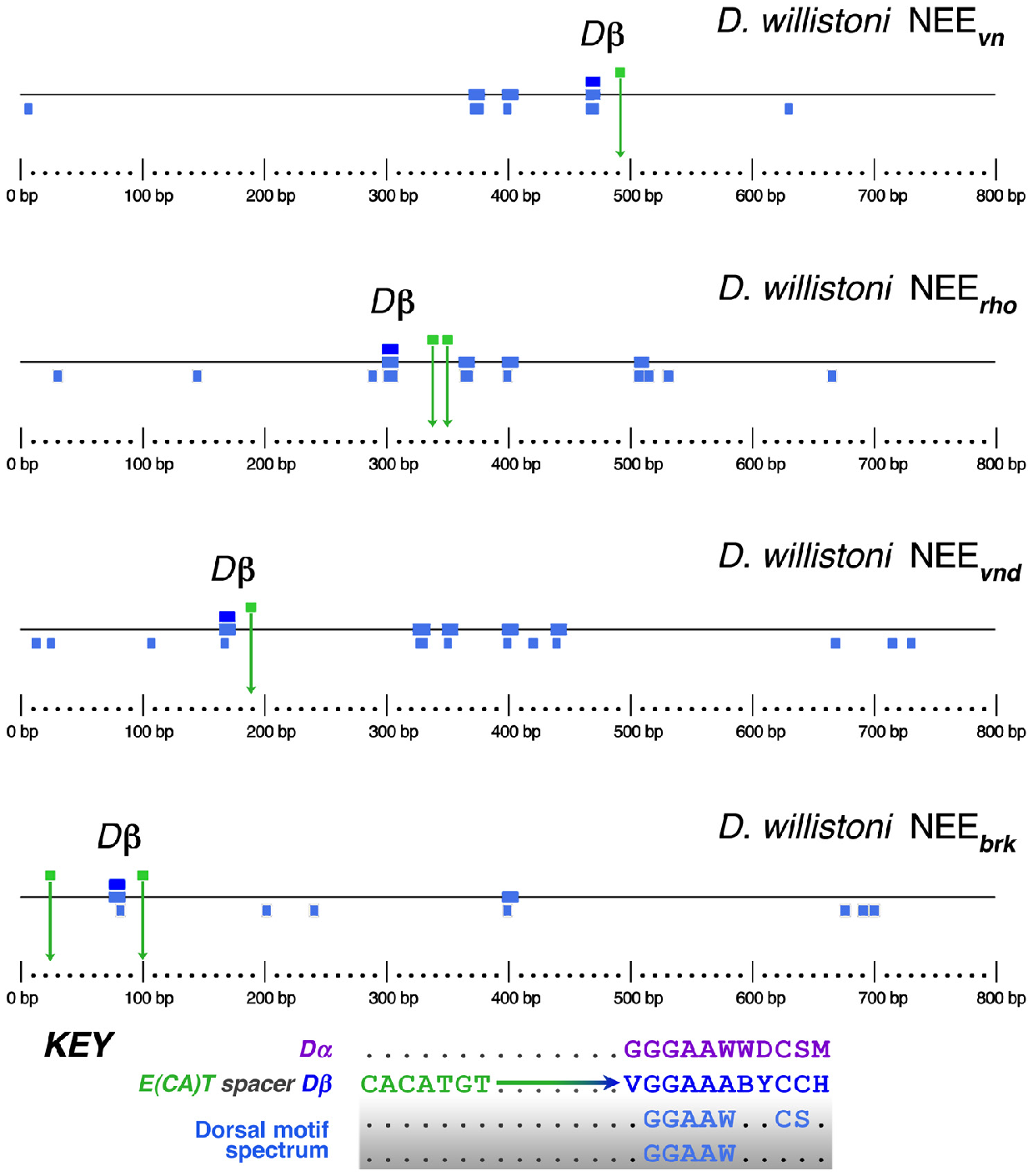}
\end{center}
\caption{
{\bf Dorsal \cis-spectra in canonical \hwil{} NEEs contain a single \hDb{} site.}\\
One Dorsal binding site variant in each cluster matches the \Db{} consensus
(dark blue).  This specialized \Db{}
site is the closest ($<$30~bp) Dorsal binding site variant to the \ECAT{}
element (green), which is a binding site for the Dorsal co-activator Twist, and
the Snail mesodermal repressor. Sites matching this specialized Dorsal binding
motif \Db{} are distinct from the \Da{} elements (numbered purple labels).
}
\label{fig6}
\end{figure}\clearpage

\begin{figure}[!ht]
\begin{center}
\includegraphics[width=5.5in]{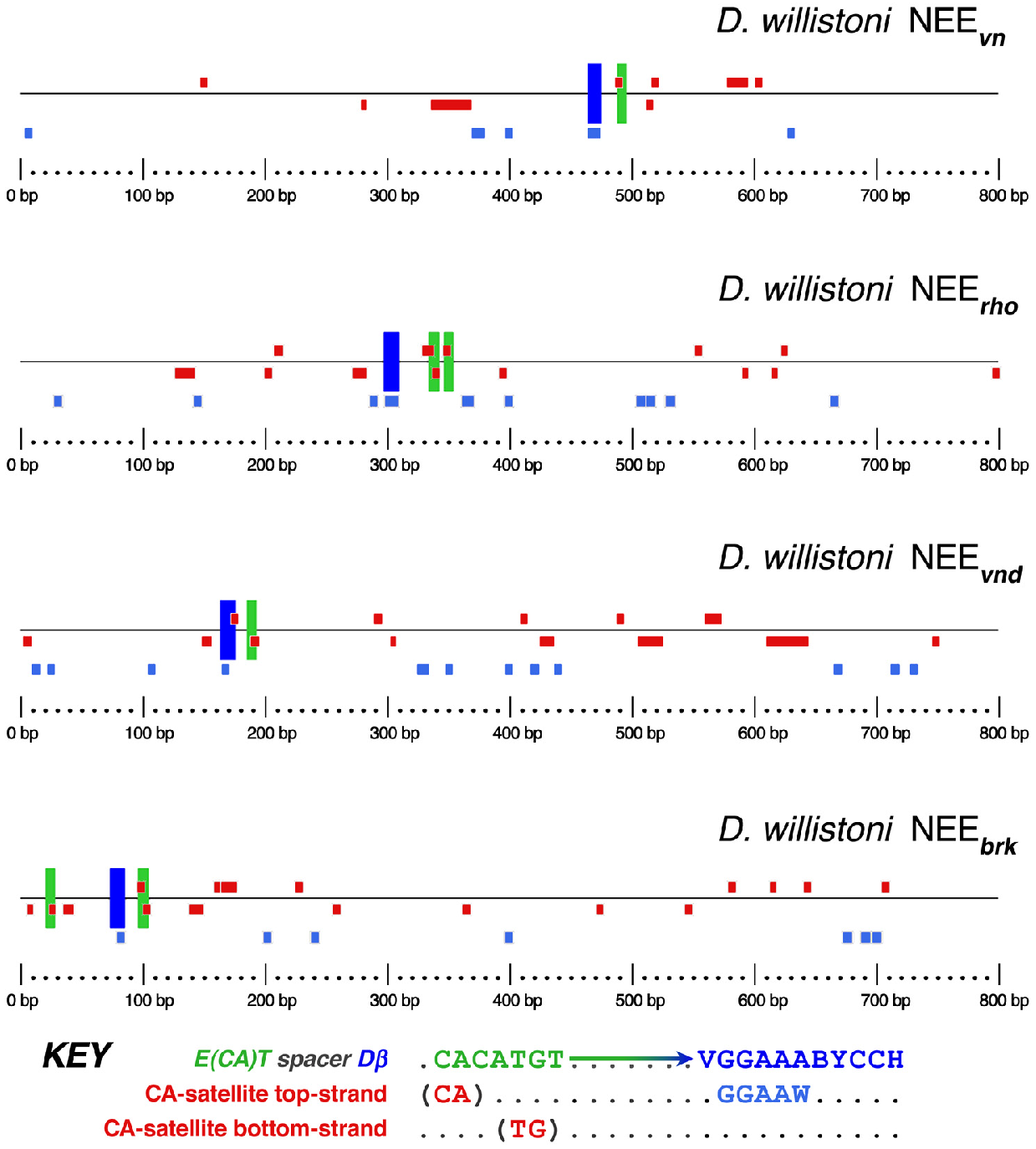}
\end{center}
\caption{
{\bf Canonical NEEs from \hwil{} are enriched with \hDNA{CA}-satellite.}\\
All canonical NEEs from the \wil{} genome also are enriched in
\DNA{CA}-satellite, almost as much as the \NEE{vnd} sequence, which was present
in the latest common ancestor of dipterans (see text). Furthermore, the longest such tracts
are associated with divergent \Db{} halfsites (pale blue).  The \NEE{vn}
\cis-spectra has expanded \DNA{CA}-satellite tracts associated with
ghost \Db{} elements at $\sim$340--400~bp and again at $\sim$580--630~bp, while
\NEE{rho} also has expanded \DNA{CA}-satellite tracts coordinated to
ghost \Db{} motifs at $\sim$130--150~bp and again at $\sim$270--290~bp. Such
signatures are consistent with the hypothesis that much of the clustering is
evidence of past deprecation events between precisely spaced \Db{} and \ECAT{}
elements. Enhancers are aligned on the unique \SUH{}/\Da{} site at position 
400~bp (see Fig.~5).
}
\label{fig7}
\end{figure}\clearpage

\begin{figure}[!ht]
\begin{center}
\includegraphics[width=5.5in]{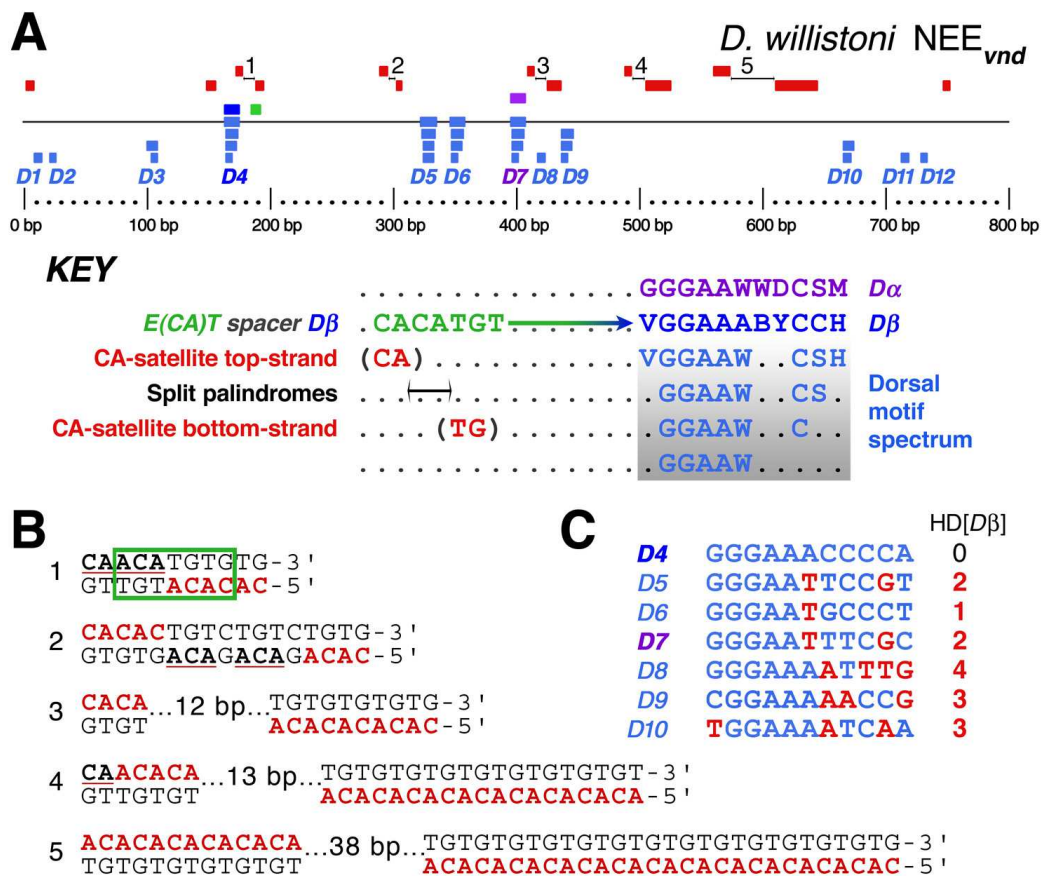}
\end{center}
\caption{
{\bf The \hvnd{} NEE from \hwil{} is enriched in split, palindromic \hDNA{CA}-satellite.}\\
Analysis of the \vnd{} NEE sequence in the relatively uncompacted \wil{} genome
indicates a long history of instability at \ECAT{} elements. Such signatures
could be variably interpreted as past selection for new \ECAT{} elements or new
optimized spacer lengths, intrinsic mutational bias for repeat expansions,
and/or both of these combined.  {\bf A)} Split, palindromic \DNA{CA}-satellite
tracts are present in the \NEE{vnd} of \wil{} as visualized by matches to short
\DNA{CA}-satellite motifs (\fiveprime{CACA} or \fiveprime{ACAC}).  The larger
palindromic \DNA{CA}-satellite tracts are numbered and their sequences shown in
B.  {\bf B)} The exact sequence composition of the \DNA{CA}-satellite indicates that
these were once intact \ECAT{} elements as found at the presumed functional
site located in palindrome \#1 (green box).  However, even the intact \ECAT{}
shows recent expansion in this lineage. Such expansions or contractions
relative to the \Db{} motif alter the precise length of the linking spacer and
consequently also alter the precise Dorsal concentration threshold of the
enhancer.  {\bf C)} Increasingly longer, and presumably older \DNA{CA}-tracts
are associated with increasingly divergent Dorsal binding site variants as
shown. For each such Dorsal binding site variant listed the Hamming Distance
(HD) or number of mismatches (red letters) from \Db{} is indicated.
}
\label{fig8}
\end{figure}\clearpage

\begin{figure}[!ht]
\begin{center}
\includegraphics[width=5.5in]{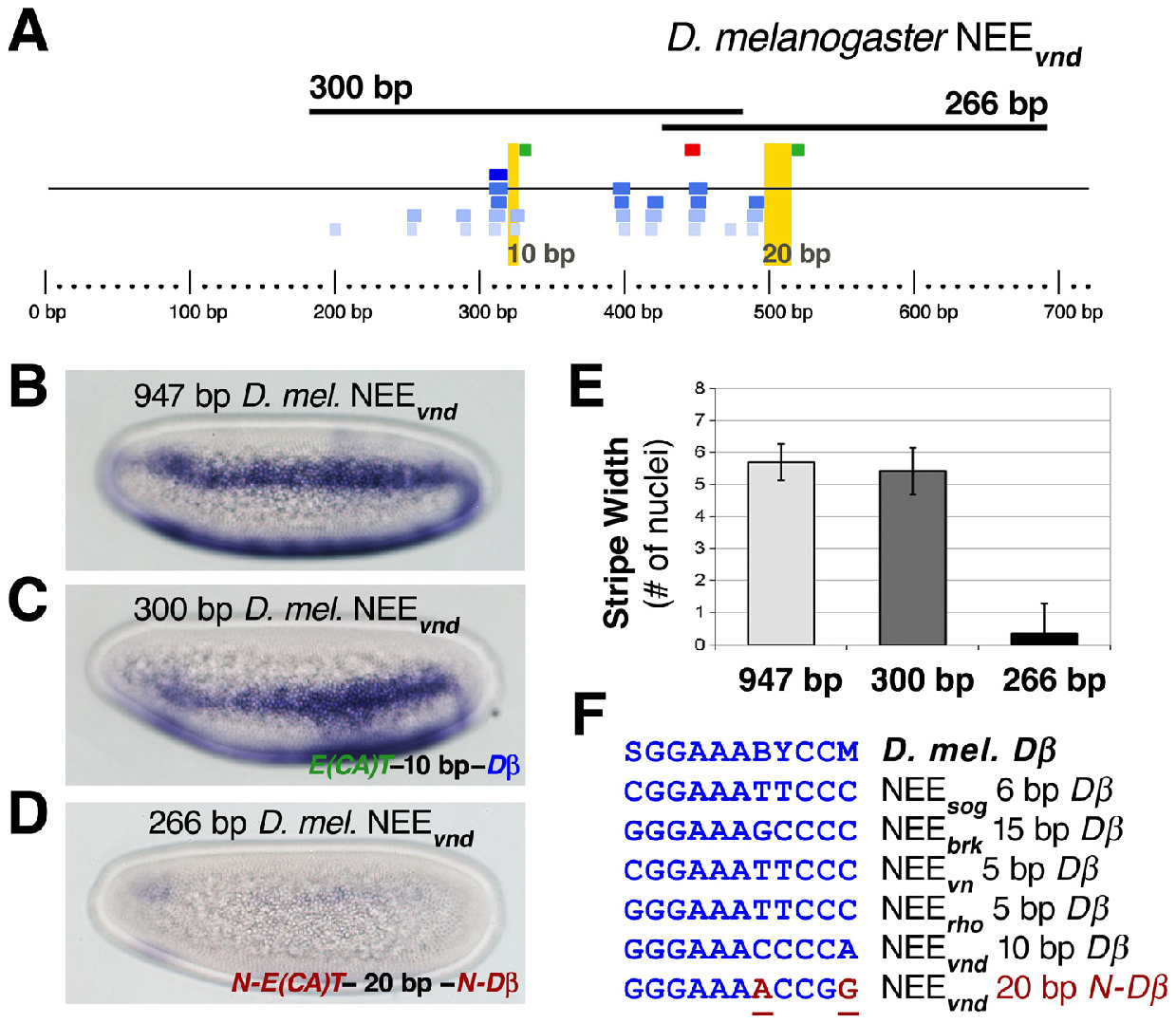}
\end{center}
\caption{
{\bf The second \hEtoD{} encoding within \hNEE{vnd} was deprecated prior to divergence of the \hDros{} genus.}\\
{\bf A)} Unlike \wil{}, the \NEE{vnd} in \mel{} has two apparently intact threshold-encodings, 
one of which is coordinated by a 10~bp spacer (narrow yellow column),
and another that is coordinated by a 20~bp spacer (wide yellow column).  Motifs
follow the key in Fig.~1 except the Dorsal binding spectra are shaded with
decreasing intensity as degeneracy increases. The 947~bp ``full-length''
fragment encompasses the entire 720~bp shown in the graphic.  Two smaller
tested fragments are shown in dark bold lines.  Both of these overlap
and include the \SUH/\Da{} site (red/blue stack).  {\bf B)} The 947~bp
\NEE{vnd} ``full-length'' enhancer sequence drives a normal pattern of {\it
lacZ} expression.  {\bf C)} The 300~bp \NEE{vnd} subfragment drives a similar
pattern as the full-length version, despite the absence of the second
coordinated Dorsal/Twist binding site pair.  {\bf D)} The 266~bp \NEE{vnd}
subfragment fails to drive a robust lateral stripe of {\it lacZ} expression at
any threshold.  Faint staining is occasionally seen in a lateral patch towards
anterior pole.  {\bf E)} Quantification of the stripe width over several
embryos for each construct depicted in A--D shows that the full-length enhancer
is not measurably different than the 300~bp fragment containing a single
\EtoD{} encoding.  {\bf F)} The Dorsal binding site coordinated by 20~bp to the
second \ECAT{} element is divergent (red letters) from the \Db{} consensus for
\mel{}.  This \mel{} \Db{} consensus matches the \Db{} consensi in other
lineages more closely than a \mel{} consensus made with the 20~bp coordinating
Dorsal binding site variant. 
}
\label{fig9}
\end{figure}\clearpage

\begin{figure}[!ht]
\begin{center}
\includegraphics[width=5.5in]{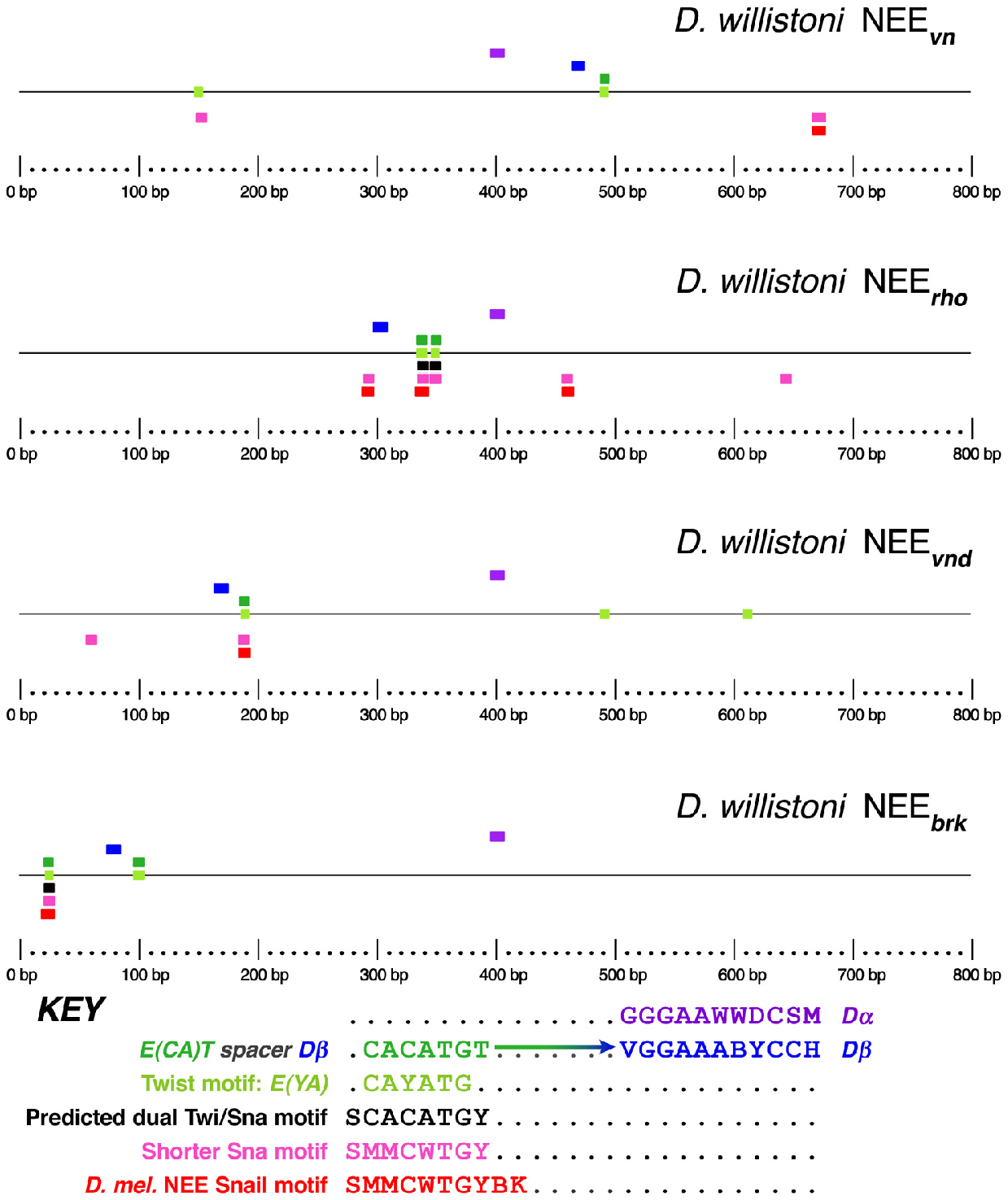}
\end{center}
\caption{
{\bf \ECAT{} versus Twist and Snail binding motifs in \hwil{} NEEs.}\\
The simple superimposition of motifs representing binding preferences for Twist bHLH
complexes and the Snail C$_2$H$_2$ zinc-finger transcriptional repressor, results in
a predicted dual motif that is similar but not identical to the observed \ECAT{} motif. 
Because the \ECAT{} motif appears to be subject to repeat expansions and contractions,
as seen in Table 2, and because this would result in threshold-modifying variants, 
we believe that the peculiar difference between the predicted dual site and the observed
invariant site, is strong support for our evolutionary model of dynamic deprecation of 
encodings via \DNA{CA}-satellite instability. These motifs are depicted here.
}
\label{fig10}
\end{figure}\clearpage

\begin{figure}[!ht]
\begin{center}
\includegraphics[width=\textwidth]{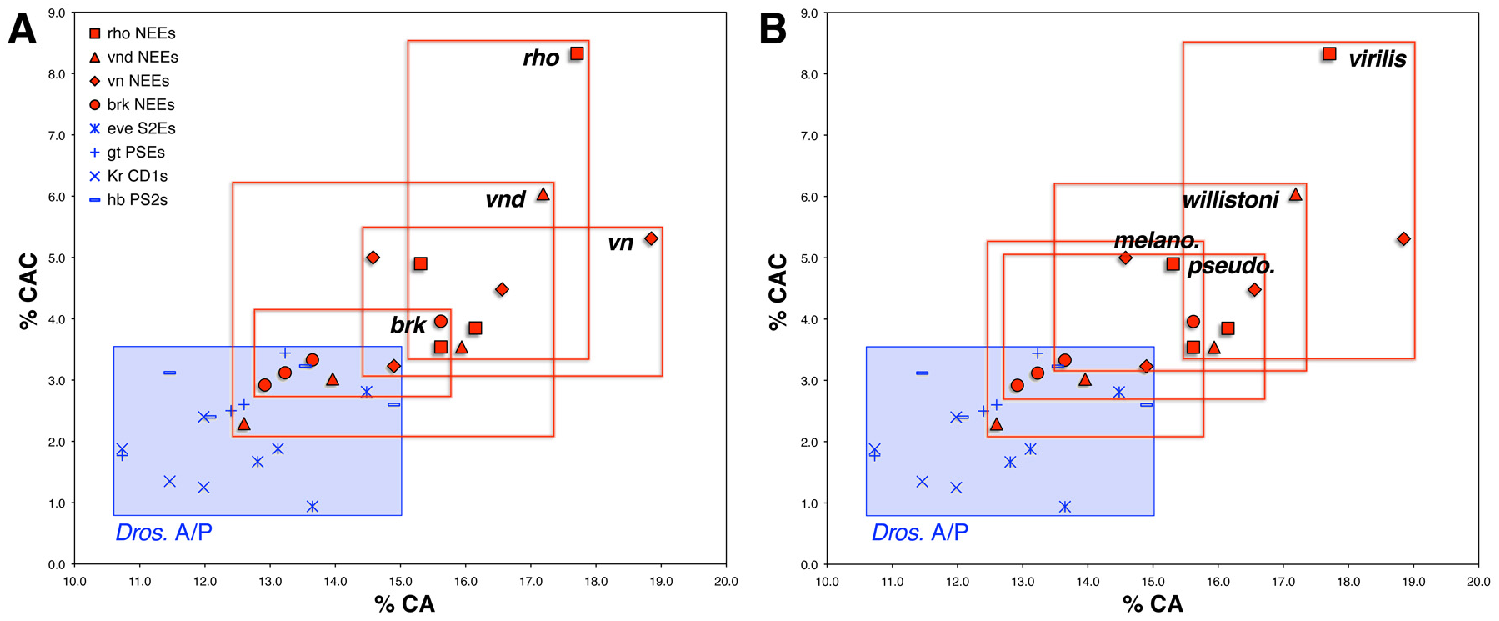}
\end{center}
\caption{
{\bf High levels of \hECAT{} fragments have accumulated in canonical NEEs across the genus.}\\
The percentage of sequence that is composed of either \fiveprime{CA}
dinucleotides or \fiveprime{CAC} trinucleotides is graphed for several
orthologous groups of enhancers from \mel{}, \pse{}, \wil{}, and \vir{}.  Each
window of NEE sequence is taken $\pm 480$~bp from \Db{} for each species. Each
window of an A/P enhancer is a 960~bp sequence centered around the Bicoid
binding site cluster. {\bf A)} Each orthologous set of NEEs is boxed separately
to visualize enrichment relative to other groups.  In contrast to the canonical
NEEs, the Bicoid binding site clusters of several canonical A/P enhancers at
the {\it eve}, {\it gt}, {\it Kr}, and {\it hb} loci are not associated with
high \DNA{CA}-satellite content. All 16 of these enhancers fit within the blue
box shown in the graph. {\bf B)} Same as A, except NEEs are boxed by species.
Because \wil{} and \vir{} represent lineages from each of the subgenera of
\Dros, this graph highlights the secondarily-derived, reduced state of
\DNA{CA}-satellite in \mel{} NEEs. 
}
\label{fig11}
\end{figure}\clearpage

\begin{figure}[!ht]
\begin{center}
\includegraphics[width=5.5in]{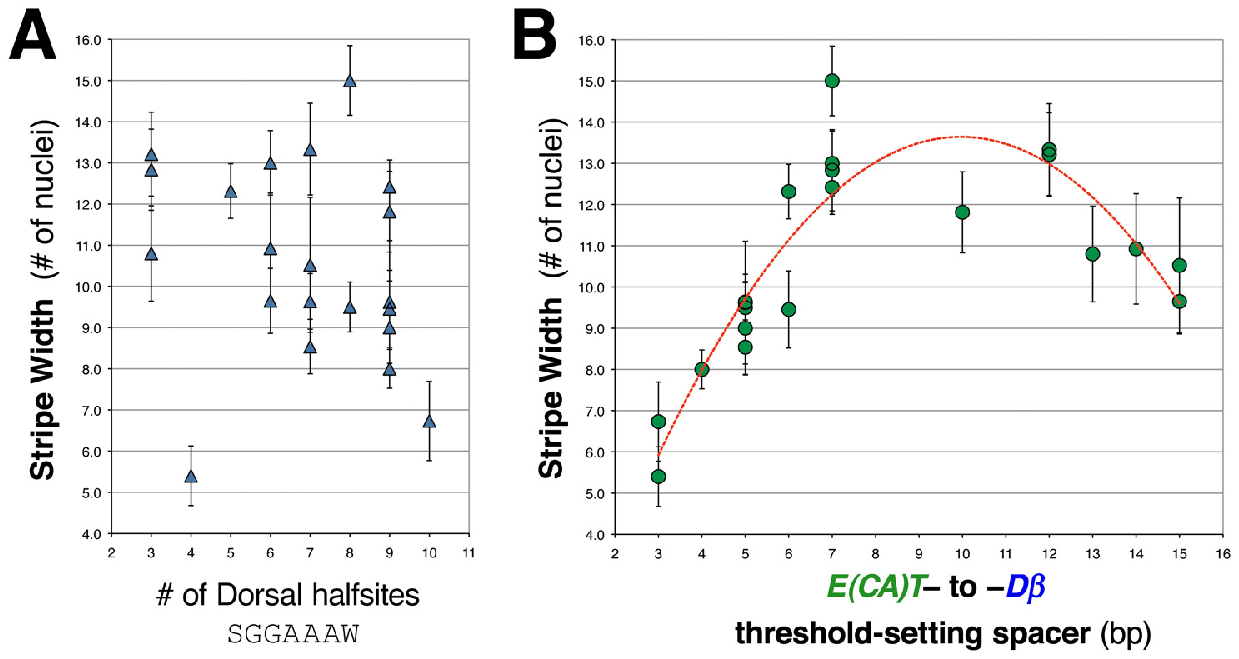}
\end{center}
\caption{
{\bf The spacer between the \hECAT{} and \hDb{} encodes the threshold-response
to the Dorsal morphogen concentration gradient, and is independent of the number or density of variant Dorsal binding sites.}\\
{\bf A)} The number of Dorsal halfsites in the $\sim$1~kb window $\pm$ 480~bp
from \Db{} from diverse NEEs of varied age, lineage, and locus, is not
predictive of the the precise Dorsal concentration threshold readout.  {\bf B)}
In contrast, the precise spacer length between the \ECAT{} and \Db{} elements
is predictive (red trendline, second order polynomial) of the precise threshold
readout over a range from 3~bp to 15~bp. Vertical axes for both graphs in A and
B are aligned for cross-referencing.
}
\label{fig12}
\end{figure}\clearpage

\begin{figure}[!ht]
\begin{center}
\includegraphics[width=3.5in]{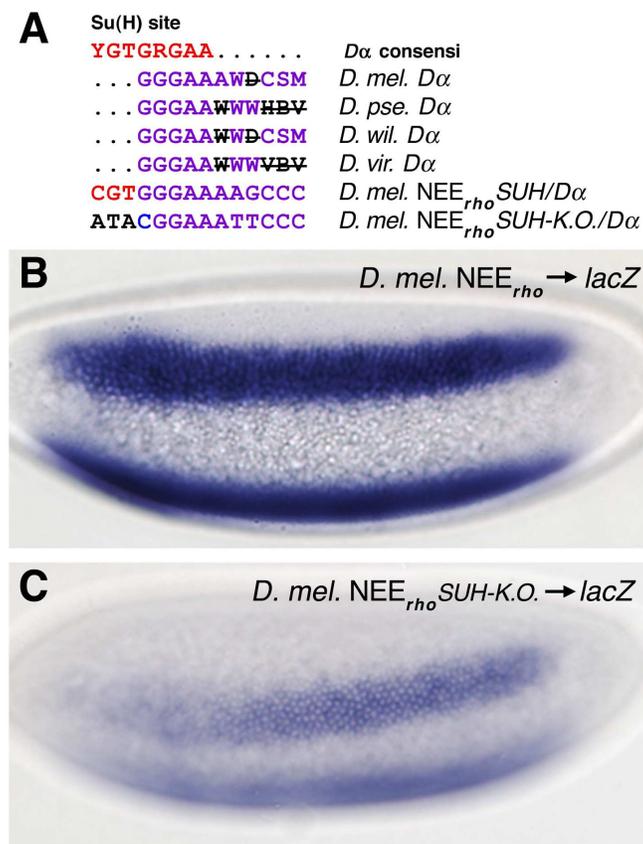}
\end{center}
\caption{
{\bf The \hDa{} motif is a \hNDb{} necro-element that was exapted into a Su(H) binding site.}\\
{\bf A)} Alignment of the lineage-specific consensi for \Da{} shows that the
portion overlapping the Su(H) binding site portion is the least divergent. The
second half of the Dorsal binding site is also increasingly degenerate (black
struck-out letters) in comparison to other lineages. Such a signature of
divergence is characteristic of drift. Based on this pattern of divergence and
the activities of more recent NEEs, we conclude that
\Da{} is non-functional and represents a deprecated \Db{} site exapted into
\SUH.  Also shown are the wild-type and mutated sequences of this site tested
in the \NEE{rho} backbone from \mel.  {\bf B--C)} Relative activities of
\NEE{rho}-driven reporters differing by the presence (B) or absence (C) of the
Su(H) binding site, via a mutation that leaves the Dorsal site intact. The
\SUH{} element is required for activity levels but not the precise Dorsal
concentration threshold encoding. This suggests that Su(H) acts after Dorsal
and Twist threshold-activation.
}
\label{fig13}
\end{figure}\clearpage

\begin{figure}[!ht]
\begin{center}
\includegraphics[width=3.5in]{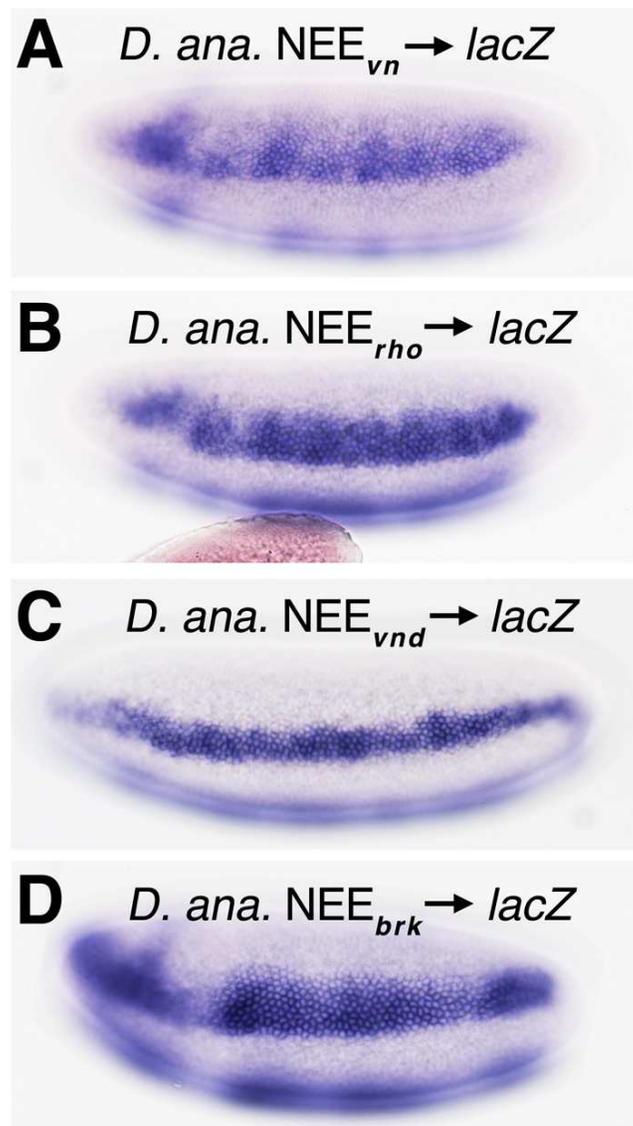}
\end{center}
\caption{
{\bf Canonical NEEs from \hana{} are functional in \hmel{} embryos.}\\
See text.
}
\label{fig14}
\end{figure}\clearpage

\begin{figure}[!ht]
\begin{center}
\includegraphics[width=5.5in]{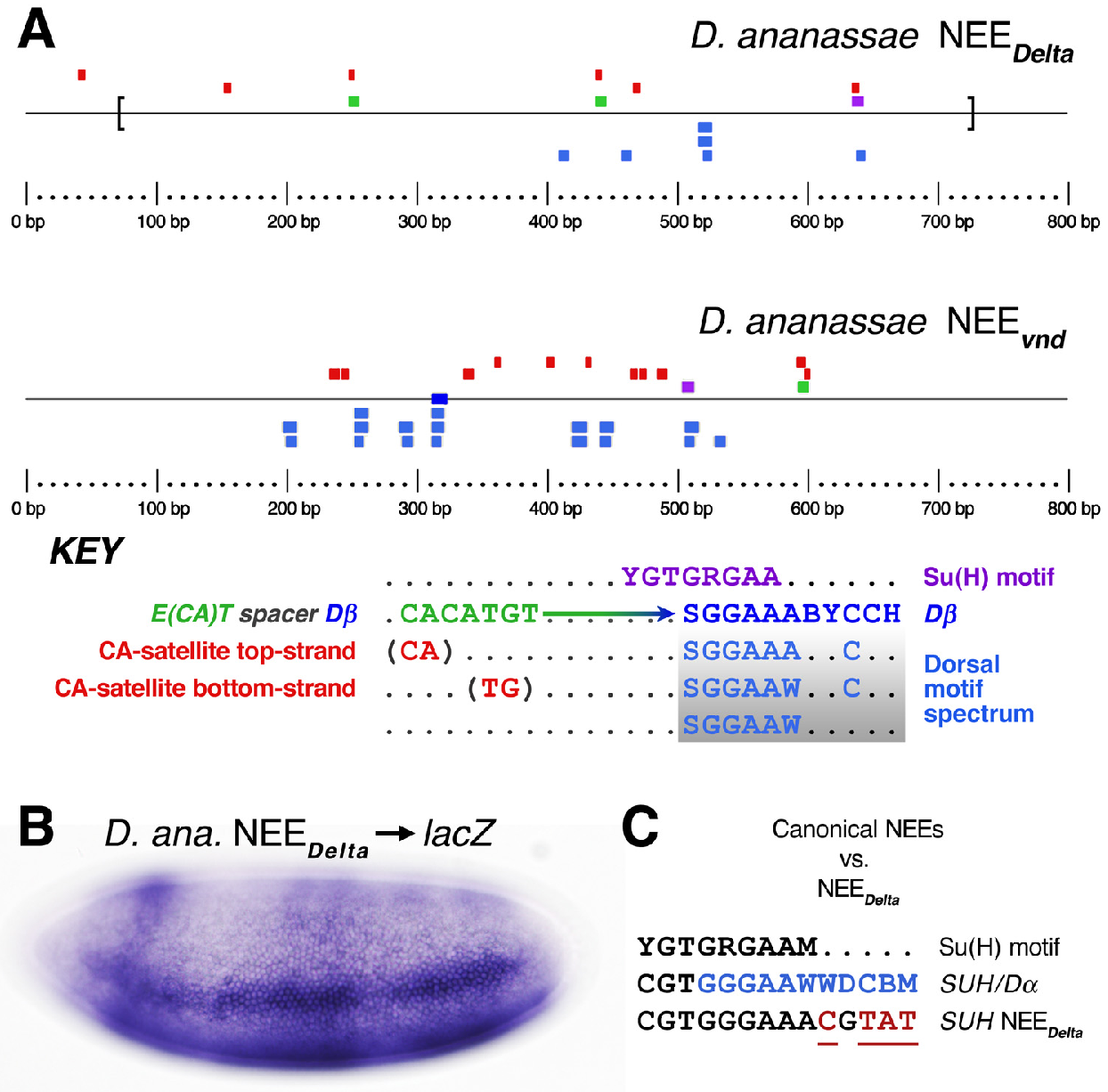}
\end{center}
\caption{
{\bf A newly evolved NEE at the \hDl{} locus of \hana{} has not yet accumulated a necro-element cluster.}\\
{\bf A)} The genome of \ana{} contains a recently-evolved enhancer \NEE{Delta}
as well as older, canonical NEEs, such as \NEE{vnd} (shown). Dorsal
\cis-spectra are associated with the canonical NEEs but not with the
\NEE{Delta} sequence, despite employing the essential NEE logic of an \EtoD{}
encoding that is near a Su(H) binding site. Brackets in the \NEE{Delta}
sequence indicate the boundaries of the fragment tested in \mel{} and shown in
B.  {\bf B)} The \NEE{Delta} module from \ana{} drives a narrow stripe of
expression spanning the $\sim$5 nuclei of the mesectoderm and ventral
neurogenic ectoderm in \mel{} embryos.  {\bf C)} The \SUH{} element does not
overlap a ghost \Da{} site. This suggests that the \SUH{} element in this
recently-evolved NEE sequence is the original site that has not yet needed to
re-evolve or track closer to the latest, functioning \EtoD{} encoding.
\DNA{CA}-satellite is defined here as sequences matching two
\DNA{CA}-dinucleotide repeats or longer (given by the UNIX regular expression:
\texttt{A?(CA)\{2,\}C?}.
}
\label{fig15}
\end{figure}\clearpage

\begin{figure}[!ht]
\begin{center}
\includegraphics[width=6.0in]{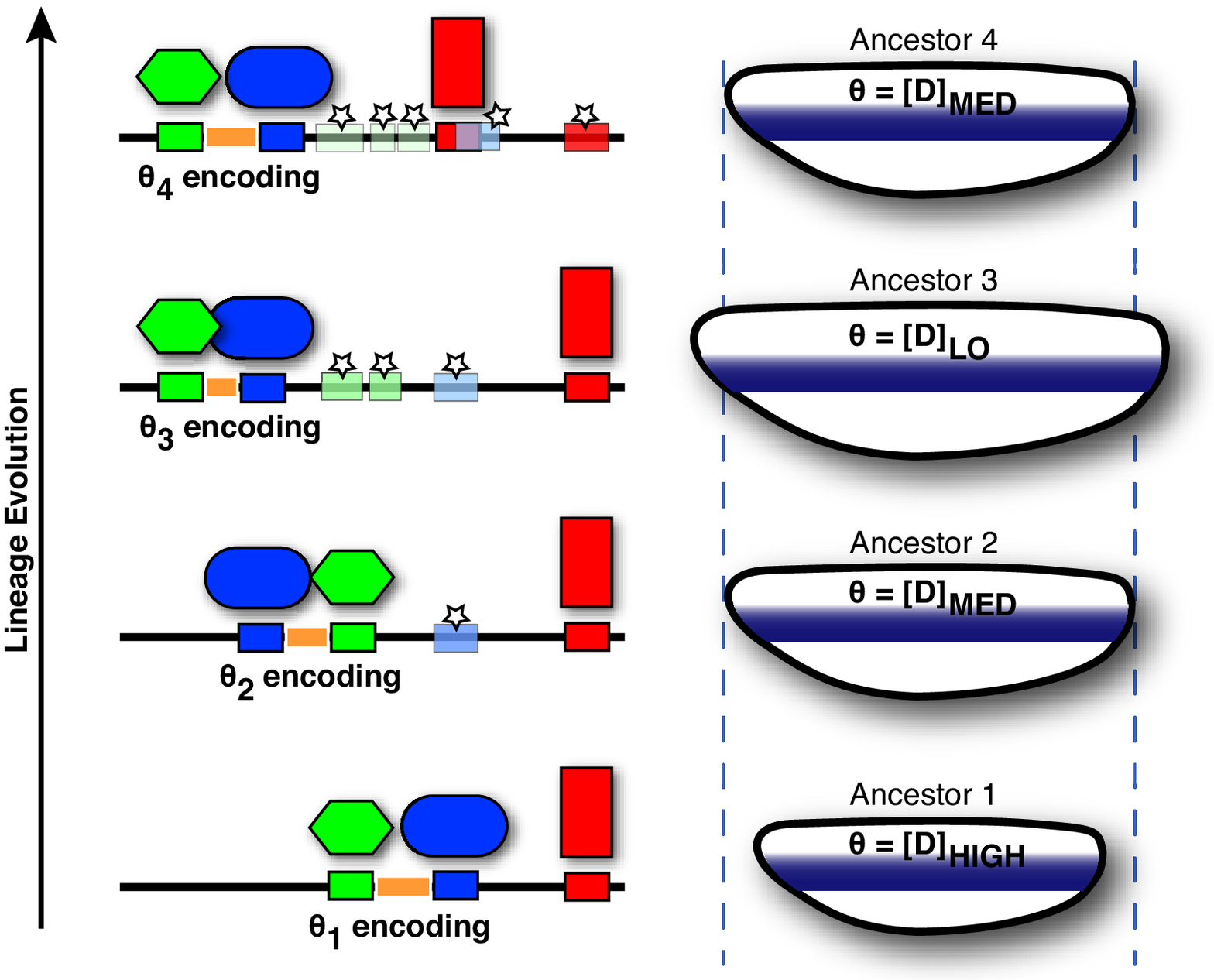}
\end{center}
\caption{
{\bf Dynamic deprecation produces necro-element clusters over time.}\\
The evolutionary maintenance of precise threshold encodings via dynamic
deprecation and re-selection of replacement encodings can be inferred for
Neurogenic Ectodermal Enhancers (NEEs). This process produces necro-element
clusters (starred, faded boxes) during the course of lineage evolution.
Depicted are binding elements for Dorsal (blue), Twist (green), and Su(H)
(red).  Spacer elements (orange) separate the Dorsal and Twist elements by a
fixed distance, whose length determines the precise threshold encoding required
for a given embryo type occurring during lineage evolution.  Because genes,
such as \vnd, which are expressed in the neurogenic ectoderm must be expressed
over the same number of cells despite evolutionary changes in the size of the
embryo (right column), selection will favor NEEs with new, compensatory,
threshold encodings (left column).  There are multiple other reasons for
selecting new threshold encodings, but these are not depicted here for
simplicity.  New encodings arise either by selection on variant spacer lengths
(\eg, evolution of threshold \#4), or by the selection of new replacement sites
defining preferred spacers (thresholds \#1--3). Su(H) sites in particular can
also be exapted from relic Dorsal necro-elements when selection favors
proximity to the current encoding (see threshold \#4) . Over time, these
processes produce a cluster of necro-elements at an enhancer.  Increasingly,
this prominent signature heavily influences future evolutionary kinetics.
}
\label{fig16}
\end{figure}\clearpage

\section*{Tables}

\begin{table}[!htpb]
\begin{flushleft}
\caption{
{\bf Specialized Dorsal motifs in \textit{\textbf{Drosophila}} NEEs.}
}
\end{flushleft}
\begin{tabular*}{\hsize}{@{\extracolsep{\fill}}llc}
\hline
Species & Motif    & Consensus over canonical NEEs  \\
\hline
\mel{} & \SUH/\Da  & \texttt{CGTGGGAAAWDC\underline{S}M} \\
\mel{} & \Db       & \texttt{\underline{NVVS}GGAAABYCCM} \\
\hline
\ana{} & \SUH/\Da  & \texttt{CGTGGGAAWWDC\underline{BM}} \\
\ana{} & \Db       & \texttt{\underline{BSVN}GGAAABYCCC} \\
\hline
\pse{} & \SUH/\Da  & \texttt{CGTGGGAA\underline{W}WW\underline{HB}V} \\
\pse{} & \Db       & \texttt{\underline{BSMS}GGAAABYCCH} \\
\hline
\wil{} & \SUH/\Da  & \texttt{YGYGGGAA\underline{W}WDC\underline{S}M} \\
\wil{} & \Db       & \texttt{\underline{DKVS}GGAAABYCC\underline{H}} \\
\hline
\vir{} & \SUH/\Da  & \texttt{CGTGGGAA\underline{W}WW\underline{VB}V} \\
\vir{} & \Db       & \texttt{\underline{KNVS}GGAAABYCCH} \\
\hline
\end{tabular*}
\begin{flushleft}
 DNA consensi for the indicated elements of canonical NEEs in each species are listed
 in IUPAC code.  Canonical NEEs are located in \vnd, \rhomb, \vn, \brk{} loci. 
 Underlined letters refer to the more degenerate site of two equivalent positions across the \Da{} and
 \Db{} consensi for that lineage. 
\end{flushleft}
\label{tab:motifs}
\end{table}
\clearpage

\begin{table}[!htpb]
\begin{flushleft}
\caption{
{\bf List of intact or nearly intact encodings in tested NEEs.}
}
\end{flushleft}
\begin{tabular*}{\hsize}{@{\extracolsep{\fill}}rrllrc}
\hline
No. &           & Enhancer                                & \ECAT{}\footnotemark[1]              & Spacer       & \Db{}\footnotemark[2]    \\
\hline
 1 &   878 bp & {\it D. mel.} \NEE{rho} wt               & \DNA{\underline{CACATGT}}            &   5 bp       & \DNA{GGGAAATTCCC}  \\
 2 &   302 bp & {\it D. mel.} \NEE{rho} wt min           & \DNA{\underline{CACATGT}}            &   5 bp       & \DNA{GGGAAATTCCC}  \\
 3 &   302 bp & {\it D. mel.} \NEE{rho} \SUH{} $\Delta$  & \DNA{\underline{CACATGT}}            &   5 bp       & \DNA{GGGAAATTCCC}  \\
 4 &   912 bp & {\it D. mel.} \NEE{vn} sp -1 bp          & \DNA{\underline{CACATGT}}            &   4 bp       & \DNA{CGGAAATTCCC}  \\
 5 &   913 bp & {\it D. mel.} \NEE{vn} wt                & \DNA{\underline{CACATGT}}            &   5 bp       & \DNA{CGGAAATTCCC}  \\
 6 &   914 bp & {\it D. mel.} \NEE{vn} sp +1 bp          & \DNA{\underline{CACATGT}}            &   6 bp       & \DNA{CGGAAATTCCC}  \\
 7 &   915 bp & {\it D. mel.} \NEE{vn} sp +2 bp          & \DNA{\underline{CACATGT}}            &   7 bp       & \DNA{CGGAAATTCCC}  \\
 8 &   918 bp & {\it D. mel.} \NEE{vn} sp +5 bp          & \DNA{\underline{CACATGT}}            &  10 bp       & \DNA{CGGAAATTCCC}  \\
 9 &   947 bp & {\it D. mel.} \NEE{vnd} wt               & \DNA{A\underline{CACATGT}}           &  10 bp       & \DNA{GGGAAACCCCA}  \\
   &          &                                          & \textit{\DNA{\underline{CACATGT}TG}}     &  \textit{20 bp}  & \DNA{GGGAAA\~{A}CCG\~{G}} \\
10 &   300 bp & {\it D. mel.} \NEE{vnd} wt trunc         & \DNA{A\underline{CACATGT}}           &  10 bp       & \DNA{GGGAAACCCCA}  \\
11 &   266 bp & {\it D. mel.} \NEE{vnd} wt trunc         & \textit{\DNA{\underline{CACATGT}TG}}     &  \~{20 bp}  & \DNA{GGGAAA\~{A}CCG\~{G}} \\
12 &   657 bp & {\it D. mel.} \NEE{brk} wt               & \DNA{CA\underline{CACATGT}GTGTTTG}   &  15 bp       & \DNA{GGGAAAGCCCC}  \\
   &          &                                          & \textit{\DNA{CAA\underline{CACATGT}T}}   &  \textit{21 bp}  & \DNA{GGGAA\~{T}GTC\~{A}A} \\
13 &   651 bp & {\it D. mel.} \NEE{brk} sp -3 bp         & \DNA{CA\underline{CACATGT}GTGTTTG}   &  12 bp       & \DNA{GGGAAAGCCCC}  \\
   &          &                                          & \textit{\DNA{CAA\underline{CACATGT}T}}   &  \textit{21 bp}  & \DNA{GGGAA\~{T}GTC\~{A}A} \\14 &   553 bp & {\it D. mel.} \NEE{sog} wt               & \DNA{C\underline{CACATGT}GT}         &  7 bp        & \DNA{CGGAAATTCCC}  \\
15 &   738 bp & {\it D. ana.} \NEE{rho} wt               & \DNA{C\underline{CACATGT}GT}         &  3 bp        & \DNA{AGGAAATTCCC}  \\
16 &   758 bp & {\it D. ana.} \NEE{vn} wt                & \DNA{\underline{CACATGT}}            &  5 bp        & \DNA{CGGAAATTCCC}  \\
17 &   642 bp & {\it D. ana.} \NEE{vnd} wt               & \DNA{CA\underline{CACATGT}T}         & 11 bp        & \DNA{GGGAAACCCCC}  \\
   &          &                                          & \textit{\DNA{\underline{CACATGT}GTTGG}}  & \textit{40 bp}   & \DNA{TGGAAA\~{AA}CC\~{G}}  \\
18 &   946 bp & {\it D. ana.} \NEE{brk} wt               & \DNA{CA\underline{CACATGT}GT$_5$GGTTTGT}   & 15 bp  & \DNA{TGGAAAGCCCC}  \\
19 &   658 bp & {\it D. ana.} \NEE{Dl} wt                & \DNA{C\underline{ACATGT}TGCTG}       &  3 bp        & \DNA{GG\~{A}AAATTCC\~{A}}  \\  
20 &   843 bp & {\it D. pse.} \NEE{rho} wt               & \DNA{\underline{CACATGT}T}           &  6 bp        & \DNA{GGGAAATTCCT}  \\  
   &          &                                          & \DNA{CC\underline{CACATGT}GTTT}      & 19 bp        & \DNA{GGGAAATTCCT}  \\  
   &          &                                          & \textit{\DNA{CCC\underline{ACATGTG}TTT}}      & \textit{45 bp}   & \DNA{CGGAAATTCCT}  \\  
21 &   858 bp & {\it D. pse.} \NEE{vn} wt                & \DNA{C\underline{CACATGT}TTGG}       &  5 bp        & \DNA{CGGAAATTCCC}  \\  
22 & 1,305 bp & {\it D. pse.} \NEE{vnd} wt               & \DNA{CA\underline{CACATGT}TGG}       & 11 bp        & \DNA{GGGAAACTCCA}  \\  
   &          &                                          & \DNA{A\underline{CACATGT}TTTT}       & 10 bp        & \DNA{GGGAA\textit{T}TCCCT}  \\  
   &          &                                          & \textit{\DNA{CA\underline{CACATGT}TGG}}       & \textit{28 bp}   & \DNA{\~{T}GGAAA\~{AA}CC\~{G}}  \\  
23 &   859 bp & {\it D. pse.} \NEE{brk} wt               & \DNA{CACAC\underline{CACATGT}GTGTTTG}& 15 bp        & \DNA{GGGAAAGCCCC}  \\  
24 &   784 bp & {\it D. wil.} \NEE{rho} wt               & \DNA{\underline{CACATGT}}            &  6 bp        & \DNA{GGGAA\~{T}TCC\~{T}A}  \\  
   &          &                                          & \DNA{CACA\underline{CACATGT}G}       & 19 bp        & \DNA{GGGAA\~{T}TCC\~{T}A}  \\  
   &          &                                          & \DNA{CACAC\underline{ACATGTG}}       & 26 bp        & \DNA{CGGAAATTCCT}  \\  
25 &   796 bp & {\it D. wil.} \NEE{vn} wt                & \DNA{ACAAAC\underline{ACATGT}}       & 14 bp        & \DNA{CGGAAATTCCC}  \\  
26 &   790 bp & {\it D. wil.} \NEE{vn} sp -7 bp          & \DNA{CAAAAC\underline{ACATGT}}       &  7 bp        & \DNA{CGGAAATTCCC}  \\  
27 &   964 bp & {\it D. wil.} \NEE{vnd} wt               & \DNA{CA\underline{CACATGT}TG}        & 11 bp        & \DNA{GGGAAACCCCA}  \\  
28 &   960 bp & {\it D. wil.} \NEE{vnd} sp +\ECAT{}      & \DNA{\underline{CACATGT}}            &  7 bp        & \DNA{CGGAAA\~{AA}CC\~{G}}  \\  
   &          &                                          & \DNA{CA\underline{CACATGT}TG}        & 11 bp        & \DNA{GGGAAACCCCA}  \\  
29 &   748 bp & {\it D. wil.} \NEE{brk} wt               & \DNA{CAA\underline{CACATGT}GTTTGGGTG}   & 13 bp     & \DNA{GGGAAAGCCCC}  \\  
30 &   742 bp & {\it D. wil.} \NEE{brk} sp -6 bp         & \DNA{CAA\underline{CACATGT}GTTT}     &  7 bp        & \DNA{GGGAAAGCCCC}  \\  
31 &   726 bp & {\it D. vir.} \NEE{rho} wt               & \DNA{C\underline{CACATGT}G}          &  7 bp        & \DNA{CGGAAATTCCT}  \\  
32 &   828 bp & {\it D. vir.} \NEE{vn} wt                & \DNA{C\underline{CACATGT}TTGTG}      &  6 bp        & \DNA{CGGAAATTCCC}  \\  
33 & 1,011 bp & {\it D. vir.} \NEE{vnd} wt               & \DNA{CA\underline{CACATGT}TG}        &  8 bp        & \DNA{GGGAAACCCCA}  \\  
34 &   756 bp & {\it D. vir.} \NEE{brk} wt               & \DNA{\underline{CACATGT}GTTTGG}      & 12 bp        & \DNA{GGGAAAGCCCC}  \\  
\hline
\end{tabular*}
\begin{flushleft}
{\footnotesize 1. \DNA{CA}-satellite extending from intact \ECAT{} elements is shown when present.  
Fragmented \DNA{CA}-satellite and their loosely coordinated Dorsal spectra are not shown. Likely deprecated encodings are italicized.}\\
{\footnotesize 2. Dorsal sites are written with the best halfsite on the top strand. \Db{} sequences departing from species' consensi are indicated with a tilde.} 
\end{flushleft}
\label{tab:NEEs}
\end{table}
\clearpage

\begin{table}[!htpb]
\begin{flushleft}
\caption{
{\bf CA-satellite content in \textit{\textbf{Drosophila}} genomes and their canonical NEE sets.}
}
\end{flushleft}
\begin{tabular*}{\hsize}{@{\extracolsep{\fill}}lrrr}
\hline
                                          & \mel              &   \wil              &  \vir             \\
                                          &  release 5.22     &   release 1.3       &  release 1.2      \\
\hline
Total DNA in assembly                     &  162,370,174~bp   &     223,610,028~bp  &   189,205,863~bp  \\
$\%$ \DNA{CA}-satellite - genome          &        $ 3.9 \%$  &          $ 4.0 \%$  &         $ 4.5 \%$ \\
$\%$ \DNA{CA}-satellite - canonical NEEs  &        $ 5.3 \%$  &          $ 7.7 \%$  &        $ 10.0 \%$ \\
$\%$ \ECAT{} - canonical NEEs             &        $ 1.3 \%$  &          $ 1.5 \%$  &         $ 1.8 \%$ \\
\hline
\end{tabular*}
\begin{flushleft}
 \DNA{CA}-satellite was defined as \DNA{CA}-dinucleotide repeats of 2 or more with an optional single
 nucleotide extension of the repeat pattern at either end.  Canonical NEE sequences for \vnd, \rhomb, \vn,
 \brk{} loci were extracted $\pm480$~bp from \Db.
\end{flushleft}
\label{tab:CA-sat}
\end{table}
\clearpage

\end{erivesformat}
\end{document}